\documentclass[12pt]{article}

\usepackage{graphicx} 
\usepackage{jheppub}
\usepackage{amsfonts,amsmath,bm} 
\usepackage{mathrsfs} 
\usepackage{setspace}
\usepackage{upgreek}
\usepackage[utf8]{inputenc} 
\usepackage{comment} 

\makeatletter
\def\@fpheader{\relax}
\makeatother

\DeclareMathAlphabet{\mathbbold}{U}{bbold}{m}{n} 
\newcommand{\be}{\begin{equation}} \newcommand{\ee}{\end{equation}}

\title{Remarks on holographic models of the Kerr-AdS$_{\mathbf{5}}$ geometry}

\author[a,b]{Juli\'{a}n Barrag\'{a}n Amado,}\emailAdd{j.j.barragan.amado@rug.nl}
\author[a]{Bruno Carneiro da Cunha,}\emailAdd{bruno.ccunha@ufpe.br}
\author[b,c]{and Elisabetta Pallante}\emailAdd{e.pallante@rug.nl}

\affiliation[a]{Departamento de F\'{i}sica, Universidade Federal de Pernambuco,
  50670-901, Recife, Pernambuco, Brazil}
\affiliation[b]{Van Swinderen Institute for Particle Physics and
  Gravity, University of Groningen, 9747 Groningen, Netherlands}
\affiliation[c]{NIKHEF, Science Park 105, 1098 XG Amsterdam,
  Netherlands}

\abstract{We study the low-temperature limit of scalar perturbations
  of the Kerr-AdS$_5$ black-hole for generic rotational parameters. We
  motivate the study by considering real-time holography of small
  black hole backgrounds. Using the isomonodromic technique, we show
  that corrections to the extremal limit can be encoded in the
  monodromy parameters of the Painlevé V transcendent, whose expansion
  is given in terms of irregular chiral conformal blocks. After
  discussing the contribution of the intermediate states to the
  quasi-normal modes, we perform a numerical analysis of the low-lying
  frequencies. We find that the fundamental mode is perturbatively
  stable at low temperatures for small black holes and that excited
  perturbations are superradiant, as expected from thermodynamical
  considerations. We close by considering the holographic
  interpretation of the unstable modes and the decaying process.}

\keywords{Black Hole Scattering, Gauge/Gravity Correspondence, Holographic
  Models.}

\begin{document}

\maketitle

\section{Introduction}

Holography has been considered a tool to study strongly-coupled
phenomena in high energy physics for some time now. In particular,
the description of strongly-coupled plasma put forward by
\cite{Policastro:2001yc,Baier:2007ix} and
\cite{Bhattacharyya:2008jc,Bhattacharyya:2008ji,Bhattacharyya:2008mz}
laid ground to a very precise procedure to study systems such as
finite temperature deformations of conformal field theories based on
${\cal N}=4$ SYM from gravitational physics. It is particularly
successful in dealing with the hydrodynamical limit of such systems,
and quantities like the entropy to viscosity ratio
\cite{Policastro:2001yc} and energy loss calculations
\cite{Gubser:2006bz,Liu:2006ug}. 

More recently, it has been stressed the necessity of modeling
rotation and vorticity into the holographic picture
\cite{McInnes:2018hid,McInnes:2020mos}, which depends on the holographic
consideration of the generic rotating, vacuum black hole in anti-de
Sitter (AdS) backgrounds -- Kerr-AdS$_5$ black hole for short
\cite{Hawking:1998kw,Hawking:1999dp}. Usually, the study relies on
thermodynamical arguments about the behavior of fluctuations in the
black hole background. For instance, in \cite{Arefeva:2020jvo} the
authors analyzed the propagation of a string in this background to
estimate energy losses in a rotating fluid. Hydrodynamical features
and the effect of rotation in large five-dimensional Anti-de Sitter
black holes were also studied in \cite{Garbiso:2020puw}.

For a given field theory, the holographic description is based on the
Ward identities associated to the conformal group currents
\cite{deHaro:2000vlm}. The departure from a given ultraviolet fixed
point in the renormalization group flow is implemented by 
the (inwards) radial evolution on the currents using Einstein
equations from the asymptotic boundary AdS structure. In
this paper, we study the validity of the description for generic
rotation parameters at low temperatures. The motivation for
this is threefold.

First and foremost, the usefulness of holography to
study the infrared (IR) behavior of the associated field theories depends on
the gravitational perturbations at low temperatures. As
we will revisit below, large black holes have a lower bound in the
temperature, so, this study will focus on small black holes in the
low temperature regime. At high temperatures, these black holes were
studied by using the hydrodynamical approach in
\cite{Bhattacharyya:2008ji}, which, among other things, recovers (flat
space) Navier-Stokes equations by assuming that the mean free path of the
components of the fluid is small when compared to the AdS scale. This
hints at the fact that the dual field theory is in the strong coupling
regime. This construction is at odds with holography, which starts
from the assumption that the field theory has small corrections to the
ultraviolet (UV) free or weakly interacting conformal point. More
specifically, the available examples of holography usually come from
theories that are weakly coupled at the UV, and become
strongly coupled at the IR. We will make an unwarranted --
although not unprecedented -- assumption that the description based on
conformal currents is still valid in the latter regime. We also remark that
it is an open question whether this generic view of holography can be
applied to the particular case of QCD-like asymptotically free
theories with a mass gap. In the best case scenario, the
holographic interpretation in the rest of this paper will refer to
generic asymptotically conformal systems, not necessarily subject to
the analysis in  \cite{Bochicchio:2017sgq}.  

Secondly, a more comprehensive study of real-time
holography \cite{Skenderis:2008dg} for the Kerr-AdS$_5$ black hole at
low temperatures is desirable. Rotation mixes time and angular
coordinates, and induces a ``non-flat'' asymptotic structure of the
black hole background. The first fact poses problems for the analytic
continuation of quantities computed in Euclidean signature and the
second gives rise to trace anomalies which are heavily dependent on
the particular field theory dual considered. We note however, that in
a ``classical limit'' where the AdS radius is much larger than the
Planck length, the quantities computed at different asymptotic
structures, in our case essentially the one arising from global AdS
$\mathbb{R}\times S^3$ and the one arising from the Poincaré patch
$\mathbb{R}^{1,3}$ are relatable by a conformal transformation. Also,
the repercussions of the AdS instabilities found in
\cite{Green:2015kur}, of which the scalar field linear unstable modes
found by the authors in \cite{Barragan-Amado:2018pxh} are a particular
example, are expected to be of fundamental relevance to the
understanding of holography, and the study presented here seemed like
a natural step further.

The third reason is somewhat technical. As argued in
\cite{Hijano:2015zsa}, black hole perturbations should be described by
a particular limit of the four-dimensional conformal block
\cite{Dolan:2003hv}. As argued in \cite{Amado:2017kao} (see also
\cite{daCunha:2016crm}), the pertinent perturbation can be computed
using \textit{two-dimensional} chiral conformal blocks. Semi-classical
aspects of the latter which are relevant to the discussion were
anticipated in \cite{Litvinov:2013sxa}, and a full fledged solution
was presented in the seminal work of \cite{Gamayun:2013auu}, also
uncovering a relation between semiclassical and $c=1$ conformal
blocks, both related to the Painlevé VI tau function.

The low temperature black hole considered here necessitates a careful
consideration of the confluence limit, which is given in terms of the
Painlevé V tau function \cite{Lisovyy:2018mnj}. Albeit in this paper
we are focused on the study of the quasi-normal modes (QNMs) for small
black holes $r_+\ll 1$, the resulting relation holds for generic
five-dimensional Kerr black holes in the near-extremal limit. On the
one hand, this is to our knowledge a new application of the Painlevé V
transcendent -- see also
\cite{daCunha:2015ana,CarneirodaCunha:2019tia}\footnote{See
  \cite{Novaes:2018fry} for the parallel story in four-dimensional
  Kerr-de Sitter black holes, using the Painlevé VI transcendent.}. On
the other hand, the  appearance of the two-dimensional conformal
blocks may be seen from one point of view as a calculational tool,
but, as we will argue, the connection between these and
four-dimensional blocks merits further investigation. 

While the writing of this work was in course,
\cite{Aminov:2020yma} came out with a similar study of
(four-dimensional) quasi-normal modes of black holes using the
properties of conformal blocks. The authors of \cite{Aminov:2020yma}
employ the usual relation between 
Fuchsian equations and semiclassical conformal blocks as the starting
point, and their relation to Seiberg-Witten theory \cite{Alday:2009aq}
has been once more studied in \cite{Nekrasov:2020qcq}. We note,
however, that there are advantages and drawbacks to each approach. The
expansion in terms of semiclassical conformal blocks is more direct and leads
directly to an asymptotic expansion of the accessory parameter (called
$K_0$ below) of the associated differential equation
\eqref{eq:standardheun}. The $c=1$  expansion used here
implements an explicit solution to the Riemann-Hilbert problem of finding
\textit{both} accessory parameters $t_0$ and $K_0$ in terms of
monodromy data. As we saw in an earlier paper
\cite{Barragan-Amado:2018pxh}, an asymptotic expansion for 
the quasi-normal modes in the five-dimensional Kerr-AdS black hole is
more effectively obtained using $c=1$ conformal blocks.

With all this in mind, we have structured the paper as follows. In
Sec. \ref{sec:kerrads5} we review the parameters of the Kerr-AdS$_5$
black hole, as well as its asymptotic structure and conserved charges
-- mass and angular momenta. We spend some time analyzing the
holographic interpretation of the asymptotic structure, in particular the
interplay between the $\mathbb{R}\times S^3$ (global structure)
and $\mathbb{R}^{1,3}$ (Poincaré patch), in a ``near the ultraviolet''
description which allows us to identify the state in the putative
field theory corresponding to the black hole and to digress about
the fate of perturbations within the scope of linear analysis.

Sec. \ref{sec:scalarperturbations}
comprises the bulk of the results. We consider scalar
perturbations of the Kerr-AdS$_5$ black hole and show that the low
temperature limit of the QNMs is given by zeros of the Painlevé V tau
function.  Using asymptotic expansions of the Painlevé V transcendent
for small parameter, we compute the QNMs frequencies for small black
holes. We then turn to discuss the CFT description of the
perturbations, and find that the contributions to the fundamental QNM
are indeed given by the vacuum conformal block, at least for small
black holes. The higher modes are correspondingly given by special
blocks with internal dimensions given by degenerate ``heavy''
operators, in a chiral version of the analysis in
\cite{Fitzpatrick:2015foa}. We use the expansions of the Painlevé tau
function and the accessory parameter to derive the expansion of the
fundamental QNM frequency for small black holes, including the first
correction for non-zero temperature and we compare the asymptotic
expression with the numerical result. Higher modes are studied
numerically in the same regime, and we find indeed the unstable modes
of \cite{Hawking:2000vt}, provided the superradiance condition is met.

In Sec. \ref{sec:decay} we discuss the holographic consequences of the
analysis, focussing on the unstable modes. We see that the naive
extrapolation of the near-ultraviolet result, pointing to the
irrelevance of the perturbation, is misguided, and that the
instabilities arising from the same perturbations as treated in
Sec. \ref{sec:scalarperturbations} may change considerably
the fate of the state as one approaches the infrared. We
also find that the spatial dependence of the decaying modes may be
confined to small angles as the mass of the state increases. We close
in Sec. \ref{sec:discussion} with some remarks on the results and
their interpretation. 

There are two Appendices. Appendix \ref{sec:adsmass} gives a self-contained
derivation of the conserved charges for pure gravity in asymptotically
AdS spaces and emphasizes the role of the conformal structure, as a
special subcase of \cite{Hollands:2005wt}. In Appendix 
\ref{sec:painleve} we review the Fredholm determinant formulation of
the Painlevé VI transcendent derived from the generic isomonodromic
tau function proposed in \cite{Gavrylenko:2016zlf}, and its dependence
on the relevant monodromy parameters is made explicit.  

\section{The Kerr-AdS$_5$ black hole}
\label{sec:kerrads5}

The Kerr-${\rm AdS_5}$ black hole is represented by the metric
\cite{Hawking:1998kw}
\begin{align}
ds^{2} =&
         -\dfrac{\Delta_{r}}{\rho^{2}}\left(dt-\dfrac{a_1\sin^{2}\theta}{1-a_1^2}d\phi_1
         -\dfrac{a_2\cos^{2}\theta}{1-a_2^2}d\phi_2\right)^{2}+\dfrac{\Delta_{\theta}
         \sin^{2}\theta}{\rho^{2}}\left(a_1\,dt-\dfrac{(r^{2}+a_1^{2})}{1-a_1^2}d\phi_1
         \right)^{2} \nonumber  \\ 
&+ \dfrac{1+r^{2}}{r^{2}\rho^{2}}\left( a_1a_2\,dt -
  \dfrac{a_2(r^{2}+a_1^{2})\sin^{2}\theta}{1-a_1^2}d\phi_1 -
  \dfrac{a_1(r^{2}+a_2^{2})\cos^{2}\theta}{1-a_2^2}d\phi_2 \right)^{2}
  \nonumber \\ 
&+ \dfrac{\Delta_{\theta}\cos^{2}\theta}{\rho^{2}}\left( a_2\,dt-
  \dfrac{(r^{2}+a_2^{2})}{1-a_2^2}d\phi_2 \right)^{2}
  +\dfrac{\rho^{2}}{\Delta_{r}}dr^{2}+
  \dfrac{\rho^{2}}{\Delta_{\theta}}d\theta^{2}, \label{eq1}   
\end{align}
where we set the AdS radius to one and
\begin{gather}
\Delta_{r} =
             \dfrac{1}{r^{2}}(r^{2}+a_1^{2})(r^{2}+a_2^{2})(1+r^2)-2M,
             \nonumber \\ 
\Delta_{\theta} = 1- a_1^2\cos^{2}\theta -
                  a_2^{2}\sin^{2}\theta, \quad\quad \ 
\rho^{2} = r^{2} + a_1^{2}\cos^{2}\theta + a_2^{2}\sin^{2}\theta,
\label{eq:bhdeltas}
\end{gather}
and $a_1$ and $a_2$ are two independent rotation parameters. The mass
relative to the ${\rm AdS}_5$ vacuum solution and the angular momenta
were the subject of some discussion
\cite{Awad:2000aj,Gibbons:2004ai,Hollands:2005wt,Olea:2006vd}. To add to the
confusion, we include our own derivation in Appendix
\ref{sec:adsmass}. The values obtained using our procedure match those
of \cite{Gibbons:2004ai}
\begin{gather}
{\cal M}=\frac{\pi
  M(2(1-a_1^2)+2(1-a_2^2)-(1-a_1^2)(1-a_2^2))}{4(1-a_1^2)^2 (1-a_2^2)^2},\\
{\cal J}_1=\frac{\pi M }{2(1-a_2^2)}\frac{a_1}{(1-a_1^2)^2},
\label{eq:asymptmass}\quad\quad
{\cal J}_2=\frac{\pi M}{2(1-a_1^2)}\frac{a_2}{(1-a_2^2)^2}.
\end{gather}

In the following it will be useful to redefine the black hole
parameters in terms of the roots of $\Delta_{r}$
\begin{equation}
  \Delta_{r}=\frac{1}{r^2}(r^2-r_0^2)(r^2-r_-^2)(r^2-r_+^2)
\end{equation}
which, for a regular, finite charged, dressed horizon black hole
satisfy $r_0^2<-1<0<r_-^2<r_+^2$. Each non-zero root $r_i$ can be
associated to a Killing horizon, although only those of $r_-$ and
$r_+$ are physical. To each horizon we can associate
the temperature $T_k$ and angular velocities $\Omega_{1,k}$ and
$\Omega_{2,k}$, $k=0,+,-$, given by
\begin{gather}
\Omega_{1,k} = \dfrac{a_1 (1-a_1^2)}{r^{2}_{k} + a_1^{2}} \qquad
  \Omega_{2,k} = \dfrac{a_2 (1-a_2^2)}{r^{2}_{k} + a_2^{2}} \nonumber \\ 
T_{k} = \dfrac{r^{2}_{k}\Delta_{r}'(r_{k})}{4\pi(r^{2}_{k} +
  a_1^{2})(r^{2}_{k} + a_2^{2})}
= \frac{r_k}{2\pi}\frac{(r_k^2-r_i^2)(r_k^2-r_j^2)}{(r_k^2+a_1^2)(r_k^2+a_2^2)},
\label{eq:bhparameters} 
\end{gather}
where the prime stands for the derivative with respect to $r$. We note
than $T_+>0$, $T_-<0$ and $T_0$ is purely imaginary.

\subsection{Asymptotic structure}

The asymptotic structure of the metric as $r\rightarrow \infty$ is
involved for non-zero rotation parameters $a_1$ and $a_2$. By the
suitable change of coordinates 
\begin{gather}
  (1-a_1^2)\bar{r}^2\sin^2\bar{\theta}=(r^2+a_1^2)\sin^2\theta,\\
  (1-a_2^2)\bar{r}^2\cos^2\bar{\theta}=(r^2+a_2^2)\cos^2\theta,\\
  \bar{t}=t,\quad\quad\bar{\phi}_1=\phi_1+a_1t,\quad\quad
  \bar{\phi}_2=\phi_2+a_2t,
\end{gather}
we arrive, after some manipulation, at the useful formulas
\begin{gather}
  1+\bar{r}^2=\frac{1-a_1^2\cos^2\theta-a_2^2\sin^2\theta}{(1-a_1^2)(1-a_2^2)}
  (1+r^2),
  \\ 
  \bar{r}^2(1-a_1^2\sin^2\bar{\theta}-a_2^2\cos^2\bar{\theta}) =
  r^2+a_1^2\sin^2\theta+a_2^2\cos^2\theta,  \\
  \frac{d\bar{r}^2}{1+\bar{r}^2}+\bar{r}^2d\bar{\theta}^2=
 \rho^2\left(
    \frac{dr^2}{\Delta_r}
  +\frac{d\theta^2}{\Delta_\theta}\right),
\end{gather}
so that, for large $\bar{r}$, the metric becomes
\begin{multline}
  ds^2=-(1+\bar{r}^2)d\bar{t}^2+\frac{d\bar{r}^2}{1+\bar{r}^2}+
  \bar{r}^2(d\bar{\theta}^2+\sin^2\bar{\theta}\,d\bar{\phi}_1^2+
  \cos^2\bar{\theta}\,d\bar{\phi}_2^2)\\ 
  +\frac{2M}{\bar{\Delta}_{\theta}^2 \bar{r}^2}
    \left(
      \frac{d\bar{r}^2}{\bar{r}^4}+\frac{1}{\bar{\Delta}_{\theta}}
      (d\bar{t}-a_1\sin^2\bar{\theta}\,d\bar{\phi}_1-
      a_2\cos^2\bar{\theta}\,d\bar{\phi}_2)^2\right) + \ldots
   \label{eq:transitionalads}
\end{multline}
where the ellipsis denotes subleading terms in $\bar{r}$ -- see the
Appendix in \cite{Arefeva:2020jvo}, and
\begin{equation}
  \bar{\Delta}_{\theta}=1-a_1^2\sin^2\bar{\theta}-a_2^2\cos^2\bar{\theta}.
\end{equation}
In the first line of \eqref{eq:transitionalads} we recognize the
metric of vacuum global AdS. 

In order to apply the procedure in Appendix \ref{sec:adsmass} to
compute the mass and angular momenta we take
$\tilde{\Omega}=\Omega=1/\sqrt{1+\bar{r}^2}$, and write
\eqref{eq:transitionalads} 
as follows
\begin{multline}
  ds^2=\frac{1}{\Omega^2}\left(-d\bar{t}^2+
    \frac{d\Omega^2}{1-\Omega^2}+(1-\Omega^2)d\Omega_3\right) +
  \\
  2M\frac{\Omega^2}{\bar{\Delta}_{\theta}^2}\left( d\Omega^2+
    \frac{1}{\bar{\Delta}_{\theta}} (d\bar{t}-a_1\sin^2
    \bar{\theta}\,d\bar{\phi}_1-
    a_2\cos^2\bar{\theta}\,d\bar{\phi}_2)^2\right) + \ldots,
  \label{eq:transasympads}
\end{multline}
where $d\Omega_3=d\bar{\theta}^2+\sin^2\bar{\theta}\,d\bar{\phi}_1^2+
  \cos^2\bar{\theta}\,d\bar{\phi}_2^2$ is the usual metric of the
3-sphere. Again, the ellipsis at the end of \eqref{eq:transasympads}
refers to subleading terms, now of order ${\cal O}(\Omega^3)$. Like in
Fefferman-Graham development \cite{Fefferman:1985aa}, the structure of
the expansion can be explained by the equations of motion. The induced
metric  $\bar{g}_{ab}$ at the conformal infinity $\Omega=0$ is given
by the first line of \eqref{eq:transasympads}, without the
$\Omega^{-2}$ factor. The $\Omega$ 
dependence of this term fixes, up to order $\Omega^4$, the conformal
structure at infinity -- see \eqref{eq:confstruct}. The term in the second line
of \eqref{eq:transasympads} defines the asymptotic charges and thus
$h_{ab}=\Omega^2\gamma_{ab}$. Its form is explained in terms of 
$n^a=\bar{g}^{ab}\nabla_b\Omega$, the unit normal to the boundary,
and the vector field 
\begin{equation}
  \chi^a=\frac{\partial}{\partial \bar{t}}+a_1\frac{\partial}{\partial
    \bar{\phi}_1}+a_2\frac{\partial}{\partial \bar{\phi}_2}=
  \frac{\partial}{\partial t}.
\end{equation}
This vector field generates an isometry of $\bar{g}_{ab}$ and of the
AdS metric. It is actually an isometry of the Kerr-AdS metric, and
the Killing vector becomes null at the horizons $r=r_0,r_-,r_+$. In
terms of the induced metric $\bar{g}_{ab}$, one has
\begin{equation}
  \bar{g}_{ab}\chi^a\chi^b=-1+(1-\Omega^2)(a_1^2\sin^2\bar{\theta}+
  a_2^2\cos^2\bar{\theta})
  =-\bar{\Delta}_{\theta}+{\cal O}(\Omega^2).
\end{equation}
We can thus write
\begin{equation}
  h_{ab}=2M\frac{\Omega^2}{\bar{\Delta}_{\theta}^2}(n_an_b+
  \frac{1}{\bar{\Delta}_{\theta}}
  \bar{\chi}_a\bar{\chi}_b)+{\cal O}(\Omega^3)
  \label{eq:firstcorrectionh}
\end{equation}
with $\bar{\chi}_a=\bar{g}_{ab}\chi^b$. The
leading contribution to the metric correction $h_{ab}$ is actually
invariant under the conformal structure chosen. Indeed, if we take
$\Omega'=\upomega \Omega$, with $\upomega\neq 0$, then the quantities
above transform as 
\begin{equation}
  n'_a=\upomega n_a+\Omega \nabla_a\omega,\quad\quad
  \bar{g}'_{ab}=\upomega^2\bar{g}_{ab},\quad\quad
  \bar{\chi}'_a=\upomega^2\bar{\chi}_{a},\quad\quad
  \bar{\Delta}'=\upomega^2\bar{\Delta}_{\theta},
  \label{eq:conformalh}
\end{equation}
and it can be checked that $h'_{ab}=h_{ab}$ in
\eqref{eq:firstcorrectionh}, up to terms of order 
$\Omega^3$. In the following, we will refer to the leading part in
\eqref{eq:firstcorrectionh} as just $h_{ab}$.  It is traceless with
respect to the pure AdS metric, and transverse
\begin{equation} 
  \nabla_ah^{ab}=0,
\end{equation}
where the covariant derivative is the one compatible with the pure AdS
metric.

The true Fefferman-Graham parameter $\varrho$, defined in
\cite{Fefferman:1985aa}, can be computed as an expansion in 
$\Omega$:
\begin{equation}
  \varrho = \Omega +\frac{1}{4}\Omega^3+\frac{1}{4}\left(\frac{1}{2}+
    \frac{M}{\bar{\Delta}_{\theta}^2}\right)\Omega^5+{\cal O}(\Omega^7),
\end{equation}
not to be confused with $\rho$ in \eqref{eq:bhdeltas}. We have the
following structure for the metric in terms of $\varrho$, following
the procedure in \cite{deHaro:2000vlm}, 
\begin{multline}
  ds^2=\frac{d\varrho^2}{\varrho^2}+\frac{1}{\varrho^2}\left(
    -d\bar{t}^2+d\Omega_3\right)
  +\frac{1}{2}\left(-d\bar{t}^2-d\Omega_3\right)\\
  +\varrho^2\left(\left(\frac{1}{16}+\frac{M}{2\bar{\Delta}_\theta^2}\right)
    (-d\bar{t}^2+d\Omega_3)+
    \frac{2M}{\bar{\Delta}_\theta^3}(d\bar{t}-a_1^2\sin^2\bar{\theta}\,
    d\bar{\phi}_1-a_2^2\cos^2\bar{\theta}\,d\bar{\phi}_2)^2\right)
  +\ldots,
  \label{eq:asymptFGdev}
\end{multline}
again up to ${\cal O}(\varrho^3)$ terms. We recognize the induced
metric $\bar{g}_{ab}$, corresponding to the induced line element
$-d\bar{t}^2+d\Omega_3$. The particular conformal structure of
$\mathbb{R}\times S^3$ is not invariant under the transformation
$\Omega\rightarrow \upomega\Omega$. By a suitable choice of $\upomega$
(see \eqref{eq:flatminkowskimetric}), one could
change the induced metric to the flat Minkowski one
$\mathbb{R}^{3,1}$. In the following, we will refer to the
$\mathbb{R}\times S^3$ induced metric as $\bar{g}_{ab}$, or
\textit{global AdS structure}, and to the flat $\mathbb{R}^{3,1}$ as
$\hat{g}_{ab}$, or \textit{flat AdS structure}.

\subsection{Holography and hydrodynamical issues}

In global AdS, the holographic stress-energy tensor can be read
directly from \eqref{eq:asymptFGdev} above and has the expression 
\cite{Arefeva:2020jvo} 
\begin{equation}
  T^{h}_{ab} = \frac{1}{4\pi G_{N,5}}\left(\frac{1}{16}\bar{g}_{ab}+
    \frac{2M}{\bar{\Delta}_\theta^2} \left( u_{(a}u_{b)}+\frac{1}{4}\bar{g}_{ab}
    \right) \right)
  \label{eq:holostressenergy}
\end{equation}
whose $M$-dependent term satisfies the conformally invariant, perfect
fluid relation with energy density $\bar{\rho}$ and pressure $\bar{P}$
\begin{equation}
  4\pi G_{N,5}\bar{\rho}=12\pi G_{N,5}\bar{P}=
  3M/2\bar{\Delta}_\theta^2,
\end{equation}
hence $\bar{P}=\bar{\rho}/3$, and the holographic stress-energy tensor is
traceless. The normalized fluid velocity is
$u^a=\chi^a/\sqrt{\bar{\Delta}_\theta}$ and it is 
divergence-free $\bar{\nabla}_au^a=0$, where $\bar{\nabla}_a$ is the
covariant derivative associated to the metric $\bar{g}_{ab}$. Despite
depicting a classical energy profile, and being dependent on the
choice of conformal structure, \eqref{eq:holostressenergy} is by
construction a valid approximation to the stress-energy tensor of the
state in the field theory in the UV limit, which holographically
corresponds to $\bar{r}\rightarrow\infty$.  

The acceleration of $u^a$ is
\begin{equation}
  u^a\bar{\nabla}_au^b=-\frac{a_1^2-a_2^2}{\bar{\Delta}_\theta}
  \sin\bar{\theta}\cos\bar{\theta}\,\frac{\partial}{\partial\bar{\theta}}
  =-\frac{1}{4\rho}\bar{\nabla}^b\rho=\frac{1}{2\bar{\Delta}_\theta}
  \bar{\nabla}^b\bar{\Delta}_\theta,
\end{equation}
and it follows that $T^h_{ab}$ in \eqref{eq:holostressenergy} is
conserved with respect to the metric $\bar{g}_{ab}$, \textit{i.e.}
$\bar{\nabla}^aT^h_{ab}=0$.  The term
independent of $M$ in \eqref{eq:holostressenergy} corresponds to the
four-dimensional gravitational conformal anomaly
\cite{deHaro:2000vlm}, and it can be disregarded for 
large enough $M$. The total energy associated to the fluid is 
\begin{equation}
  4\pi G_{N,5} E = 4\pi G_{N,5}\int_{\mathrm{S}^3}d\Omega_3\, \bar{\rho}
  = \frac{3\pi^2M}{(1-a_1^2)(1-a_2^2)},
\end{equation}
which can be compared with the total ADM mass ${\cal M}$ in
\eqref{eq:asymptmass}. 

The perfect fluid form of the holographic stress-energy tensor for the
Kerr-AdS black hole in \eqref{eq:holostressenergy} mirrors the result
for four-dimensional backgrounds from \cite{Cardoso:2013pza}, 
and is in fact expected on general grounds. The conformally invariant
nature of \eqref{eq:holostressenergy} hints at an underlying
conformally invariant field theory dual. Following the holographic
renormalization group flow towards the IR, the corrections to the
metric will in general break this conformal invariance, which is
interpreted as the departure from a conformal fixed point in the field
theory. If we think of the field theory as free or weakly coupled at
the UV, the flow towards the IR may bring the theory to a strongly
coupled regime, resulting in, among other things, corrections to the 
perfect fluid form of its stress-energy tensor
\eqref{eq:holostressenergy}. In the spirit of holographic
renormalization group flow, we will assume that the radial evolution
given by Einstein equations will give the proper picture for the dual
field theory as the energy scale is brought down.

For high-temperatures, one can interpret the results of
\cite{Bhattacharyya:2007vs} and \cite{Bhattacharyya:2008ji}
as the equivalence between the general relativity equations and the
Navier-Stokes equations for the fluid mechanics. In our study, we want
to consider low temperatures, outside the scope of their
analysis. This necessarily leads us to consider small 
black holes, because large black holes in AdS are hot. To see this
latter point, let us recall that, when written in terms of $r_+$,
$a_1$ and $a_2$, the temperature of the black hole outer horizon
\eqref{eq:bhparameters} is 
\begin{equation}
  2\pi T_+ =  r_+\left(\frac{r_+^2+1}{r_+^2+a_1^2}+
    \frac{r_+^2+1}{r_+^2+a_2^2}-\frac{1}{r_+^2}\right),
  \label{eq:outerhortemp}
\end{equation}
which, for constant $r_+$, is a monotonically decreasing function of
$a_1$ and $a_2$. However, for $a_1,a_2<1$, $T_+$ is always larger than
$T_{\mathrm{min}}=(2r_+^2-1)/(2\pi r_+)$. We can therefore conclude
that there cannot be a zero-temperature black hole for $r_+^2>1/2$.

There are basically two ways of dealing with this issue. One is to
forgo our misgivings and simply take the calculations we will be
performing as definitions of the field theory system in the infrared
regime for small $T_+$. The other is to view the correspondence
between the gravitational and the field theoretical system as
dynamical, and give up on the equilibrium view of the fluid, from
which the equivalence between mean free path and $r_+$ was
deduced. Either choice will have interesting consequences. 

Dwelling in the second choice for a moment, if the fluid is no longer
in equilibrium, there is no reason to restrict the analysis to
time-independent configurations. The conformal transformation to
conformally flat boundary coordinates mixes time and space and induces
time-dependence for the dynamics of the state.  Using the (classical)
conformal symmetry of the fluid, let us then perform the
transformation of $\bar{g}_{ab}$ to conformally flat coordinates. The
transformation 
\begin{equation}
  \hat{t} =
  \frac{1}{2}\tan\frac{\bar{t}+\bar{\chi}}{2}+
  \frac{1}{2}\tan\frac{\bar{t}-\bar{\chi}}{2},\qquad 
  \hat{r} =
  \frac{1}{2}\tan\frac{\bar{t}+\bar{\chi}}{2}-
  \frac{1}{2}\tan\frac{\bar{t}-\bar{\chi}}{2},
\end{equation}
where
\begin{equation}
  \cos\bar{\chi} = \sin\bar{\theta}\cos\bar{\phi}_1,\qquad
  \tan\hat{\theta}=\frac{\cos\bar{\theta}}{\sin\bar{\theta}\sin\bar{\phi}_1}
  \label{eq:flatcoordchange}
\end{equation}
and $\hat{\phi}=\bar{\phi}_2$, turns the metric into
\begin{equation}
  d\hat{s}^2=\frac{-d\hat{t}^2+d\hat{r}^2+\hat{r}^2(d\hat{\theta}^2+
    \sin^2\hat{\theta}\,d\hat{\phi}^2)}{(1+\frac{1}{4}(\hat{t}+\hat{r})^2)
    (1+\frac{1}{4}(\hat{t}-\hat{r})^2)}.
  \label{eq:flatminkowskimetric}
\end{equation}
We recall that we use bars to refer to coordinates in the
$\mathbb{R}\times S^3$ manifold
($\{\bar{t},\bar{\theta},\bar{\phi}_1,\bar{\phi}_2\}$) and hatted
coordinates for Minkowski space $\mathbb{R}^{1,3}$
($\{\hat{t},\hat{r},\hat{\theta},\hat{\phi}\}$). The flat metric will
be henceforth referred to as $\hat{g}_{ab}$.

We can use the conformal symmetry of the leading order perturbation
$h_{ab}$ in \eqref{eq:conformalh} to derive the Poincaré coordinates,
flat AdS version of the holographic stress-energy tensor, in the same
approximation as \eqref{eq:holostressenergy}. The perturbation is
explicitly given by the same formula as \eqref{eq:firstcorrectionh}
\begin{equation}
  h_{ab}=2M\frac{z^2}{\hat{\Delta}^2}(n_an_b+
  \frac{1}{\hat{\Delta}}\hat{\chi}_a\hat{\chi}_b)
\end{equation}
where $n_a=(dz)_a$, $\hat{\chi}_a=\hat{g}_{ab}\chi^b$, and
\begin{multline}
  \hat{\Delta}=-\hat{g}_{ab}\chi^a\chi^b \\ =
  (1+\tfrac{1}{4}(\hat{t}+\hat{r})^2)(1+\tfrac{1}{4}(\hat{t}-\hat{r})^2)-
  (1+\tfrac{1}{4}(\hat{t}^2-\hat{r}^2))^2a_1^2-
  \hat{r}^2(a_1^2\cos^2\hat{\theta}+a_2^2\sin^2\hat{\theta}).
  \label{eq:flatdelta}
\end{multline}
The flat-space version of the stress-energy tensor
$\hat{T}^h_{ab}=\upomega^{-2}T^h_{ab}$, with
\begin{equation}
  \upomega^2 = (1+\tfrac{1}{4}(\hat{t}+\hat{r})^2)
  (1+\tfrac{1}{4}(\hat{t}-\hat{r})^2)
\end{equation}
is conserved with respect to the
flat metric $\hat{g}_{ab}$ (see Appendix D in \cite{Wald:1984}), and it
reads
\begin{equation}
  \hat{T}^h_{ab}=\hat{\rho} \hat{u}_a
  \hat{u}_b+\hat{P}(\hat{g}_{ab}+\hat{u}_a\hat{u}_b),\qquad
  \hat{\rho} = 3\hat{P}=\frac{1}{4\pi
    G_{N,5}}\frac{3M}{2\hat{\Delta}^2},
  \label{eq:flatstressenergy}
\end{equation}
and $\hat{u}^a=\chi^a/\sqrt{\hat{\Delta}}$ is the normalized velocity
of the fluid with respect to $\hat{g}_{ab}$. 

\begin{center}
  \begin{figure}[tb]
  \includegraphics[width=0.325\textwidth]{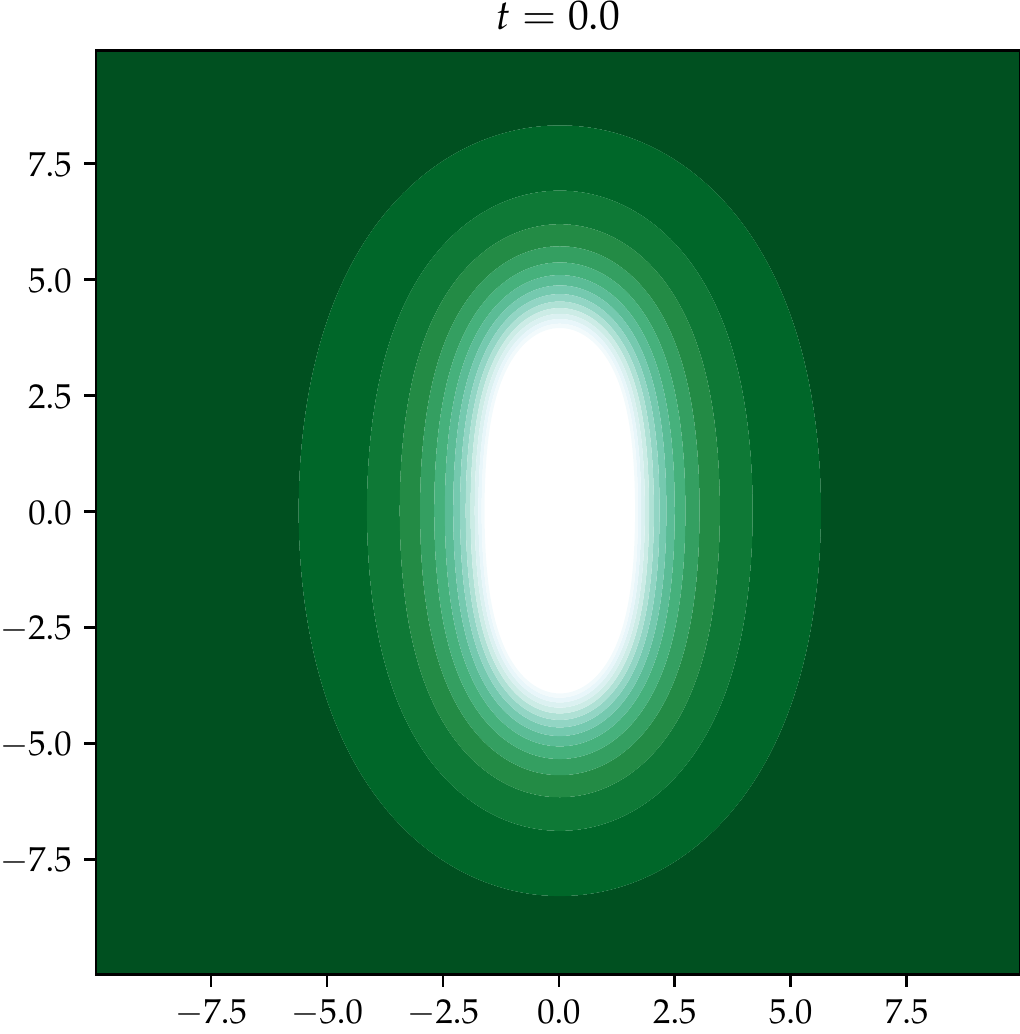}
  \includegraphics[width=0.325\textwidth]{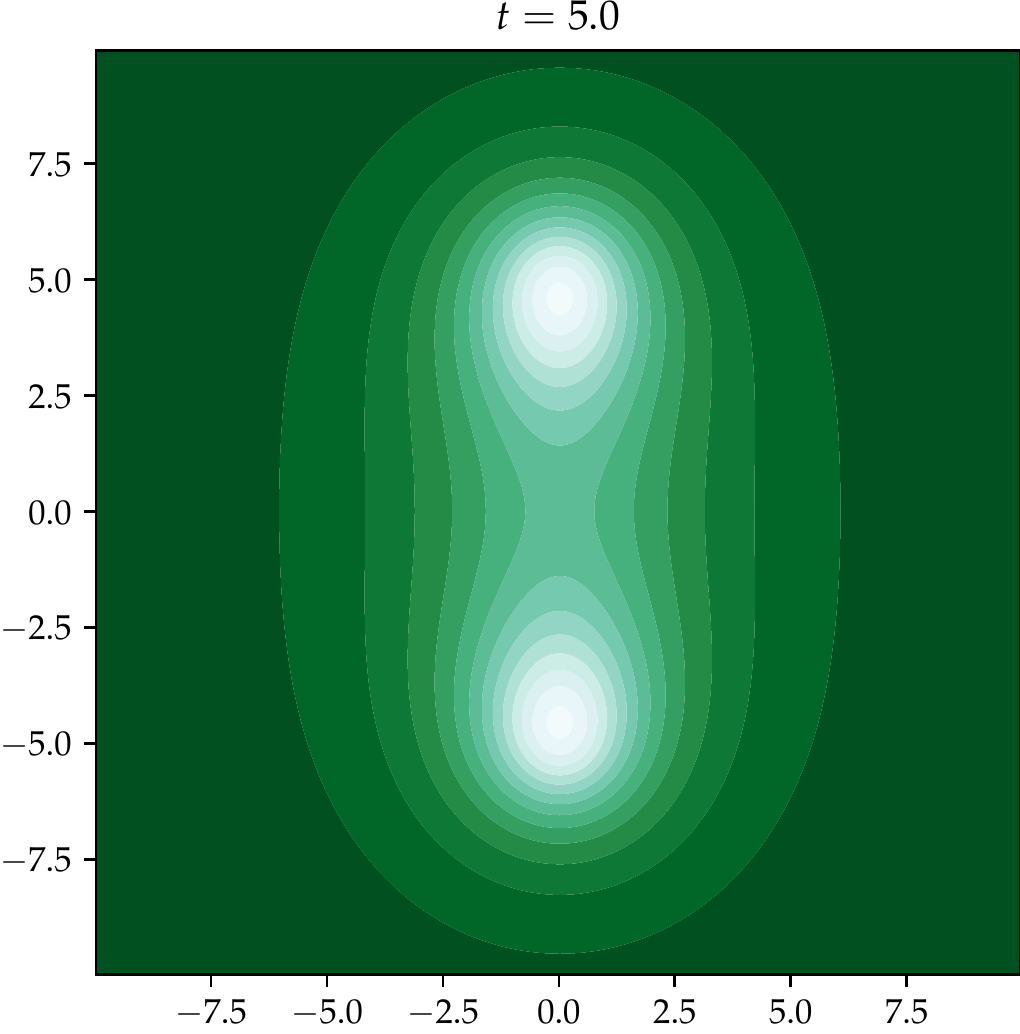}
  \includegraphics[width=0.325\textwidth]{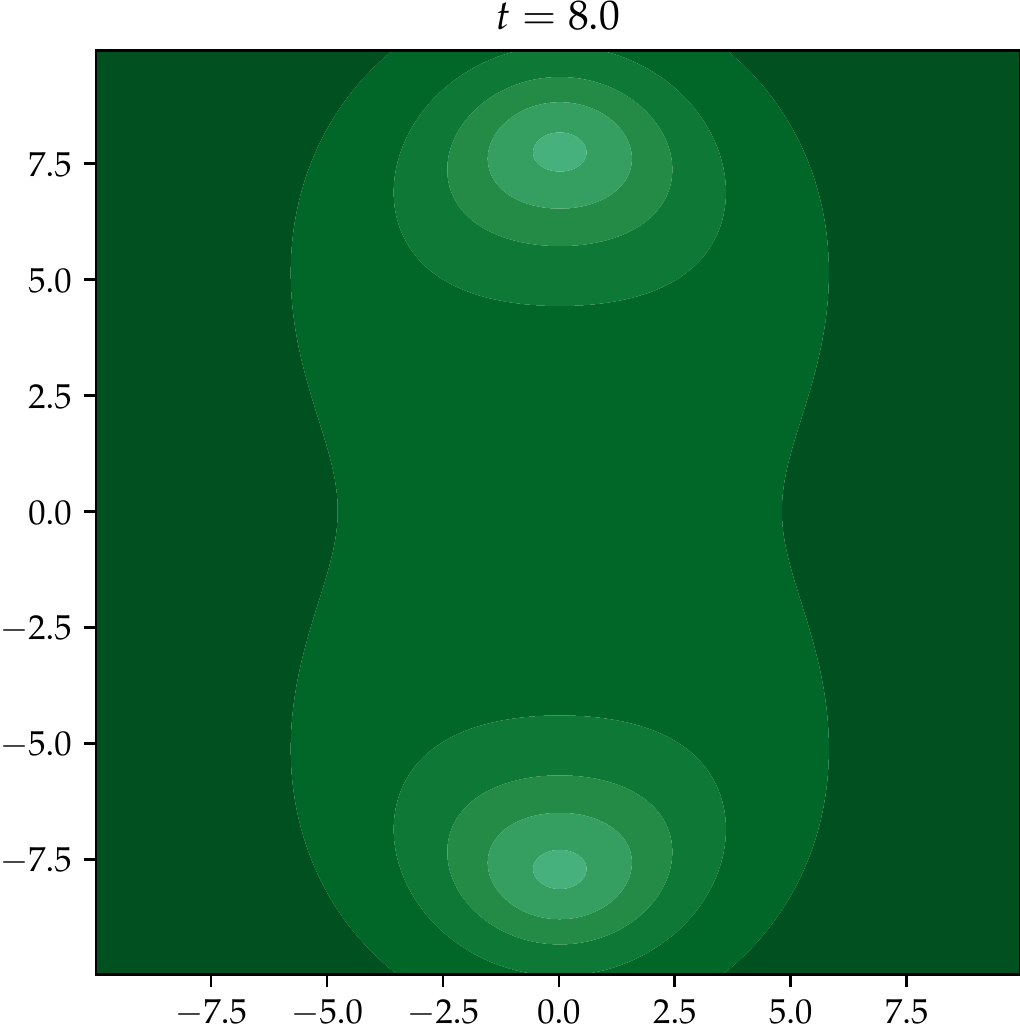}
\caption{Three snapshots of the energy density profile
  \eqref{eq:flatstressenergy}, as sagittal projection. The free
  propagation shows two ``blobs''  of radiation traveling along the
  $z$ axis and meeting close to the origin as $t\rightarrow 0$. The
  two parameters $a_1$ and $a_2$ control the spread of each blob away
  from the symmetry axis and the astigmatism as
  the two blobs meet.}
\end{figure}
\end{center}

In order to study the near ultraviolet behavior in the presence of the
perturbation, we turn to the effect of the asymptotic form of the
correction $h_{ab}$. The pure AdS metric encodes the conformal
structure -- at least of its conformal part. One can then interpret
the effect of $h_{ab}$ as encoding the departure from the ultraviolet
fixed point as one turns on the renormalization group flow towards the 
infrared.

One should note at this point that there is an ambiguity in choosing
$\hat{\phi}=\bar{\phi}_2$ in passing from the $\mathbb{R}\times S^3$
conformal structure to Minkowski space $\mathbb{R}^{3,1}$. The
Kerr-AdS black hole background can be interpreted as the resulting CFT
(thermal) state obtained by turning expectation values for the
operators comprising the generators of the Cartan subalgebra of the
conformal group $\mathrm{SO}(4,2)$ 
\begin{equation}
  \frac{\partial}{\partial \bar{t}} \sim M_{01},\qquad
  \frac{\partial}{\partial \bar{\phi}_1} \sim M_{23},\qquad
  \frac{\partial}{\partial \bar{\phi}_2} \sim M_{45}
  \label{eq:so42charges}
\end{equation}
where the right-hand side makes use of an explicit representation of
the conformal group as six-dimensional ``Lorentz'' generators of
$\mathbb{R}^{2,4}$. In choosing $\hat{\phi}=\bar{\phi}_2$, we
explicitly chose $M_{45}$ as the angular momentum generator in the
four-dimensional picture. One could as well choose $M_{23}$, which
would amount to setting $\hat{\phi}=\bar{\phi}_1$. The effect of this
choice would be simply to interchange $a_1$ with $a_2$ in quantities
such as \eqref{eq:flatdelta} and \eqref{eq:flatstressenergy}. We will
come back to this point in Sec. \ref{sec:decay}.

A comparatively simple and preliminary calculation is to study the
effect on the conformal dimension $\Delta$, as read from the two-point
function of two scalar primary operators as we depart from the
conformal point in the ultraviolet. In general, one expects a number
of these primary operators to share the same conformal dimension, and
the presence of the background does not lift this degeneracy. As it is
well-known, the primary operators satisfy a conformal Casimir Ward
identity, which can be seen as the scalar wave equation in AdS. To
give the calculation a purpose independent of the 
background conformal structure chosen at infinity, we will introduce
the $\mathrm{SO}(4,2)$ invariant quantity $\sigma$: 
\begin{equation}
  \sigma = \frac{z_i^2+z_j^2+|{\bf x}_i -{\bf x}_j|^2}{2z_iz_j},
\end{equation}
which is related to the invariant $\mathrm{AdS}$ geodesic
distance $\ell$ by $\sigma = \cosh \ell$.  As it is well-known, the
pure AdS scalar propagator
\begin{equation}
  G(\sigma)=\langle \Phi(z_i,{\bf x}_i)\Phi(z_j,{\bf x}_j)\rangle
  =2^{\Delta}K\sigma^{-\Delta}{_2F_1}(\tfrac{1}{2}\Delta,
  \tfrac{1}{2}\Delta+\tfrac{1}{2};\Delta-1;\sigma^{-2}),
  \label{eq:vacuumg}
\end{equation}
reproduces the conformal two-point function for two local primary
operators with conformal dimension $\Delta$ sitting at the same value
of the cut-off parameter $z=z_i=z_j$. As one considers large values of
$z$, however, the simple CFT expression $|{\bf x}_i-{\bf
  x}_j|^{-2\Delta}$ no longer holds, since keeping the coordinate $z$
fixed (and large) is akin to keeping fixed the length of the flux tube
between the insertions \cite{Polyakov:2000jg}, which is not usually
achieved in QFT calculations. At any rate, as seen in
\cite{Ferrara:1973vz} -- see also \cite{Amado:2017kao}, this
dependence can be trivialized for pure AdS by a suitable choice of the
$z$ coordinate. 

The correction
$h_{ab}$ to the pure AdS metric $g_{ab}$ at spatial infinity
modifies this dependence (indices are raised with the vacuum metric):
\begin{equation}
  \bar{\nabla}^2G=\nabla^2G-{g}^{ab}\,{g}^{cd}(
  {\nabla}_ah_{db}-{\nabla}_dh_{ab}){\nabla}_cG
    -h^{ab}\,{\nabla}_a\,{\nabla}_bG = \Delta(\Delta-4)G,
\label{eq:gcorrection}
\end{equation}
where ${g}_{ab}$ and ${\nabla}_a$ are the vacuum AdS metric and
covariant derivative, and we are assuming non-coincident insertion
points. The last term $\Delta(\Delta-4)$ is equal to the AdS mass
squared, which is equal to the $\mathrm{SO}(4,2)$ quadratic
Casimir. Continuing with the calculation of \eqref{eq:gcorrection}, we
have, due to the tracelessness of $h_{ab}$,  
\begin{equation}
  g^{ab}g^{cd}(\nabla_ah_{db}-\nabla_dh_{ab})=\nabla_ah^{ac},
\end{equation}
where indices are raised with the pure $\mathrm{AdS}_5$ metric
$g^{ab}$. By means of the formulas \eqref{eq:tildeconnection} and
\eqref{eq:tildelevicivita}, with the choice of $\tilde{\Omega}=z$ so that 
$\tilde{g}_{ab}=\hat{g}_{ab}$ in Appendix A, we relate this divergence to
\begin{equation}
  \nabla_ah^{ac}=z^4\left(\hat{\nabla}_a\hat{h}^{ac}-
    \frac{4M z}{\hat{\Delta}^2}\hat{n}^c\right)=0,
\end{equation}
which vanishes owing to the tracelessness of $h_{ab}$ and the relation
between the acceleration of $u^a$ and the gradient of
$\hat{\Delta}$. The term $h^{ab}\nabla_a\nabla_bG$ in
\eqref{eq:gcorrection} gives less trouble, and the resulting equation is
\begin{equation}
  \left(\sigma^2-1-z^4\hat{h}^{ab}\nabla_a\sigma\nabla_b\sigma
  \right)\frac{d^2G}{d\sigma^2}+5\sigma\frac{dG}{d\sigma}
  =\Delta(\Delta-4)G,
\end{equation}
where $\hat{h}^{ab}=\hat{g}^{ac}h_{cd}\hat{g}^{db}$ has its indices
raised by the flat metric $\hat{g}_{ab}$. Using the formula for
$\sigma$ given in terms of the flat coordinates and $z$,
\begin{equation}
  z^4\hat{h}^{ab}\nabla_a\sigma\nabla_b\sigma=
  \frac{2M}{\hat{\Delta}^2(\mathbf{x}_i)}|\mathbf{x}_{ij}|^4+{\cal
    O}(z^2,|\mathbf{x}_{ij}|^6),
\end{equation}
the simple Ansatz $G\propto \sigma^{-\Delta'}$ then captures the
asymptotic behavior of the solution, with
\begin{equation}
  \begin{aligned}
  \Delta' &=\Delta+\frac{\Delta(\Delta+1)}{2(\Delta-2)}
  \left(\frac{z^4}{|\mathbf{x}_{ij}|^4}+
    \frac{2M z^4}{\hat{\Delta}^2(\mathbf{x}_i)}\right)+\ldots\\
  &= \Delta_{\mathrm{AdS}}\left(1+\frac{1}{4\pi
    G_{N,5}}\frac{2(\Delta+1)}{3(\Delta-2)}
  \hat{\rho}z^4\right)+\ldots
\end{aligned}
\label{eq:deltacorrections}
\end{equation}
where we can read the anomalous dimension as the correction to the
pure AdS value $\Delta_{\mathrm{AdS}}$, defined by the expansion of
the vacuum propagator \eqref{eq:vacuumg}, that is, in the absence of
the black hole.

Note that the effect of the background, at least perturbatively, is to
\textit{increase} the conformal dimension of the perturbation for
$\Delta>2$, which applies in particular to the marginal case
$\Delta=4$. For $z$ infinitesimally close to zero and $\Delta$
sufficiently close to $4$, \eqref{eq:deltacorrections} tells us that,
as the correction to the AdS value $\Delta_{\mathrm{AdS}}$ increases,
$\Delta'$ crosses the marginal value of $4$ and the corresponding
perturbation becomes irrelevant, and as such should not alter
substantially the fate of the flow as one approaches the infrared
limit. On the other hand, the fact that corrections to the bare AdS
value do become large at $z^4\sim M^{-1}$ can be seen as hint that the 
theory is becoming increasingly strongly coupled towards the IR. As we 
will see in the following, the remark about the fluctuations not
substantially altering the fate of the state as it flows will have to
be revised.

\section{Scalar perturbations}
\label{sec:scalarperturbations}

A complete holographic interpretation of a scalar perturbation
requires the analysis of the field in the exact Kerr-AdS$_5$
background, which we will consider in the low temperature regime.
Let us start with generic considerations on the scalar field. The 
Klein-Gordon equation in Kerr-AdS$_5$ is separable in terms of two 
(Fuchsian) ordinary differential equations, which can be cast 
in the standard form:
\begin{equation}
  \begin{gathered}
    \frac{d^2y}{dz^2}+p(z)\frac{dy}{dz}+q(z)y(z) =0,\\
    p(z) =
    \frac{1-\theta_0}{z}+\frac{1-\theta_t}{z-t_0}+\frac{1-\theta_1}{z-1},\\
    q(z) = \frac{(\theta_0+\theta_t+\theta_1+\theta_\infty-2)
      (\theta_0+\theta_t+\theta_1-\theta_\infty-2)}{4z(z-1)}-
    \frac{t_0(t_0-1)K_0}{z(z-1)(z-t_0)},
  \end{gathered}
  \label{eq:standardheun}
\end{equation}
where the set $\{\theta_0,\theta_t,\theta_1,\theta_\infty\}$ are
called the \textit{single monodromy parameters} and $\{t_0,K_0\}$ are called the
\textit{accessory parameters}. For the Klein-Gordon field mode
$\Phi_{n,\ell,m_1,m_2}$, with
\begin{equation}
  \Phi_{n,\ell,m_1,m_2}(t,r,\theta,\phi_1,\phi_2)=e^{-i\omega t+i m_1\phi_1+i
    m_2\phi_2}R_{n,\ell,m_1,m_2}(r)S_{\ell,m_1,m_2}(\theta),
\end{equation}
the radial system's single monodromy parameters are
$$
  \theta_{\mathrm{Rad},0}=\theta_-,\qquad
  \theta_{\mathrm{Rad},t}=\theta_+,\qquad
  \theta_{\mathrm{Rad},1}=2-\Delta,\qquad
  \theta_{\mathrm{Rad},\infty}=\theta_0
$$
where  
\begin{equation}
  \theta_{k} = \dfrac{i}{2\pi}\left(\dfrac{\omega -
    m_{1}\Omega_{k,1} - m_{2}\Omega_{k,2}}{T_{k}}\right), \quad
  k=\pm,0,\qquad
  \theta_{\infty}=2-\Delta,
\label{eq:radialmonodromies}
\end{equation}
and the respective accessory parameters are
\begin{equation}
  t_{\mathrm{Rad},0}=z_0=\frac{r_+^2-r_-^2}{r_+^2-r_0^2},
  \label{eq:radialt0}
\end{equation}
\begin{multline}
  4z_0(z_0-1)K_{\mathrm{Rad},0}=
  -\frac{C_{\ell}+r_{-}^2\Delta(\Delta-4)-\omega^2}{r_+^2-r_{0}^2}
-(z_0-1)[(\theta_{-}+\theta_{+}-1)^2-\theta_{0}^2-1]\\
-z_0\left[2(\theta_{+}-1)(1-\Delta)+(2-\Delta)^2-2\right].
\label{eq:accessorykradial}
\end{multline} 

The angular system's single monodromy parameters are given by
\begin{gather}
  \theta_{\mathrm{Ang},0}=m_1,\qquad
  \theta_{\mathrm{Ang},t}=m_2,\qquad
  \theta_{\mathrm{Ang},1}=2-\Delta,\qquad
  \theta_{\mathrm{Ang},\infty}\equiv\varsigma=\omega+a_1m_1+a_2m_2,
\label{eq:angmonodromies}
\end{gather}
and the accessory parameters are
\begin{equation}
  t_{\mathrm{Ang},0}=u_{0}=\frac{a_2^2-a_1^2}{a_2^2-1},
  \label{eq:angulart0}
\end{equation}
\begin{multline}
4u_0(u_0-1)K_{\mathrm{Ang},0}
=-\frac{C_{\ell}-\omega^2-a_1^2\Delta(\Delta-4)}{1-a_2^2}-u_0\left[(m_2+
  \Delta-1)^2-m_2^2-1\right] \\ 
-(u_0-1)\left[(m_1+m_2+1)^2 -\varsigma^2 -1\right].
 \label{eq:accessoryqangular}
\end{multline}

In previous work \cite{Amado:2017kao,Barragan-Amado:2018pxh}, the
authors stressed how the eigenvalue problem for the angular equation
and the QNM problem can be solved in terms of the
\textit{composite monodromy parameters} associated with the
differential equation \eqref{eq:standardheun}. In particular, the
transformation between the accessory parameters $\{t_0,K_0\}$ and the
composite monodromy parameters $\{\sigma_{0t},\sigma_{t1}\}$ is
defined in terms of the Painlevé VI transcendent tau function:
\begin{gather}
  \tau(\theta_0,\theta_t,\theta_1,\theta_\infty;
  \sigma_{0t},\sigma_{t1};t_0)=0,\\
  \frac{\partial}{\partial
    t}\log\tau(\theta_0,\theta_t-1,\theta_1,\theta_{\infty}+1;
  \sigma_{0t}-1,\sigma_{t1}-1;t_0)-
  \frac{(\theta_t-1)\theta_1}{2(t_0-1)}-\frac{(\theta_t-1)\theta_0}{2t_0}=K_0.
  \label{eq:monoconditions}
\end{gather}
The expansion of the tau function for small $t_0$ is given in Appendix
\ref{sec:painleve}, and the reader is referred to
\cite{Gavrylenko:2016zlf} for further details.

As derived in \cite{Amado:2017kao}, the quantization condition for the angular
eigenfunctions can be cast in terms of the monodromy parameters:
\begin{equation}
  \sigma_{\mathrm{Ang},0t}=-\ell, \qquad \ell\in \mathbb{N},
  \label{eq:sigmaell}
\end{equation}
from which we can compute the small $u_0$ expansion of the separation
constant $C_\ell$:
\begin{multline} \label{eq:lambda_ell}
C_{\ell} = \omega^2+\ell(\ell+2)-\varsigma^2-\frac{a_{1}^{2}+a_{2}^{2}}{2}\left(\ell(\ell+2)-\varsigma^{2}-\Delta(\Delta-4)\right) \\
-\frac{\left(a_{1}^{2}-a_{2}^{2}\right)
  \left(m_{1}^{2}-m_{2}^{2}\right)}{2\ell(\ell+2)}
\left(\ell(\ell+2)-\varsigma^{2} 
+(\Delta-2)^2 \right) \\
-\frac{(a_{1}^{2}-a_{2}^{2})^{2}}{1-a^{2}_{2}}\bigg[\frac{\left(\ell(\ell+2)+m_{2}^2-m_1^2\right)\left(\ell(\ell+2)+(\Delta-2)^2-\varsigma^2\right)}{2\ell(\ell+2)}
\\ -\frac{13}{32}\ell(\ell+2)+\frac{1}{32}\left(5+14(m^2_{1}+\varsigma^{2})-18(m^{2}_{2}+(\Delta-2)^{2})\right) \\
-\frac{\left((m_{1}+1)^{2}-m_2^{2}\right)\left((1-m_1)^{2}-m^{2}_{2}\right)\left((\Delta-1)^{2}-\varsigma^{2}\right)\left((\Delta-3)^{2}-\varsigma^{2}\right)}{32(\ell-1)(\ell+3)}
\\
+\frac{\left((m^{2}_{1}-m^{2}_{2})((\Delta-2)^{2}-\varsigma^{2})+8\right)^{2}-64-2(m^{2}_{1}+m^{2}_{2})((\Delta-2)^{2}-\varsigma^{2})^{2}}{32\ell(\ell+2)} \\
-\frac{2(m^{2}_{1}-m^{2}_{2})^{2}((\Delta-2)^{2}+\varsigma^{2})}{32\ell(\ell+2)}
-\frac{\left(m^{2}_{1}-m^{2}_{2}\right)^{2}\left((\Delta-2)^{2}-\varsigma^{2}\right)^{2}}{64}\left(\frac{1}{(\ell+2)^{3}}-\frac{1}{\ell^{3}}\right)\bigg] \\
+ \mathcal{O}\left(\left(\frac{a_{1}^{2}-a_{2}^{2}}{1-a_{2}^2}\right)^{3}\right).
\end{multline}

The radial QNMs can also be determined with similar
techniques. As seen in \cite{Barragan-Amado:2018pxh}, the purely
ingoing boundary condition at infinity 
requires that the composite monodromy parameter satisfies:
\begin{equation}
  R_{n,\ell,m_1,m_2}(r)\sim
  \begin{cases}
    (r-r_+)^{\frac{1}{2}\theta_+},\quad r\rightarrow r_+;\\
    r^{2-\Delta},\quad r\rightarrow \infty,
  \end{cases}
  \quad\Longrightarrow\quad
  \cos\pi\sigma_{1t}=\cos\pi(\theta_t+\theta_1).
  \label{eq:sigman}
\end{equation}
This condition can be conveniently written in terms of the expansion
parameter $s$ that enters the tau function expansion \eqref{eq:tildes}. From
the explicit representation for the monodromy matrices (see
also \cite{Its:2016jkt}), we have 
\begin{multline}
  \sin^2\pi\sigma_{0t}\cos\pi\sigma_{1t} =
  \cos\pi\theta_0\cos\pi\theta_\infty+
  \cos\pi\theta_t\cos\pi\theta_1\\
  -\cos\pi\sigma_{0t}(\cos\pi\theta_0\cos\pi\theta_1+
  \cos\pi\theta_t\cos\pi\theta_\infty)\\
  -\frac{1}{2}(\cos\pi\theta_\infty-\cos\pi(\theta_1-\sigma_{0t}))
  (\cos\pi\theta_0-\cos\pi(\theta_t-\sigma_{0t}))s\\  
  -\frac{1}{2}(\cos\pi\theta_\infty-\cos\pi(\theta_1+\sigma_{0t}))
  (\cos\pi\theta_0-\cos\pi(\theta_t+\sigma_{0t}))
  s^{-1}.  
  \label{eq:sigma1t}
\end{multline}
Note that this equation is of the form $A-Bs-Cs^{-1}=0$, and by direct
calculation it can be checked that
$$
A=B+C+\sin^2\pi\sigma_{0t}(\cos\pi(\theta_t+\theta_1)
-\cos\pi\sigma_{1t}).
$$
Then, due to the quantization condition \eqref{eq:sigman}, the roots
are readily seen to be $s=1$ and $s=C/B$, with the last one compatible
with the parametrization of the monodromy matrices. The parameter $s$
can be written explicitly in terms of the monodromy parameters as
\begin{equation}
  s=\frac{\sin\frac{\pi}{2}(\theta_t+\theta_0+\sigma)
    \sin\frac{\pi}{2}(\theta_t-\theta_0+\sigma)}{
    \sin\frac{\pi}{2}(\theta_t+\theta_0-\sigma)
    \sin\frac{\pi}{2}(\theta_t-\theta_0-\sigma)}
  \frac{\sin\frac{\pi}{2}(\theta_1+\theta_\infty+\sigma)
    \sin\frac{\pi}{2}(\theta_1-\theta_\infty+\sigma)}{
    \sin\frac{\pi}{2}(\theta_1+\theta_\infty-\sigma)
    \sin\frac{\pi}{2}(\theta_1-\theta_\infty-\sigma)}.
  \label{eq:radialconditions}
\end{equation}
This expression for $s$, when substituted in the tau function with the
$\theta_k$ parameters of the radial equation, will allow us through
\eqref{eq:monoconditions} to constrain the frequencies $\omega$, and
thus determine the QNMs of the scalar perturbation. 

\subsection{The low-temperature limit}

At small $r_+$, one can check from \eqref{eq:outerhortemp} that the
requirement of positive temperature limits the range of available
$a_1$ and $a_2$, which will be taken to be of order $r_+$. Bearing
this in mind, we define the parameters 
\begin{equation}
  \begin{gathered}
  \epsilon=\frac{r_+^2-r_-^2}{2r_+^2},\qquad
  \alpha_1=\frac{a_1}{r_+},\qquad
  \alpha_2=\frac{a_2}{r_+},\qquad
  \alpha^2_+ = \frac{a_1^2+a_2^2}{2r_+^2},\quad\quad
  \alpha^2_- = \frac{a_1^2-a_2^2}{2r_+^2},
  \end{gathered}
\end{equation}
with the understanding that $\alpha_+$ will be of order $1$. We will
see below that the value of $\epsilon$ will be constrained by
temperature requirements, whereas $\alpha_-$ is completely determined
from $\epsilon$, $\alpha_+$ and $r_+$.

The radii and mass parameters are in turn given in terms of $\epsilon$, and
$\alpha_\pm$ by  
\begin{gather}
  r_-^2=(1-2\epsilon) r_+^2,\quad\quad
  -r_0^2=1+2(1+\alpha^2_+-\epsilon)r_+^2,\\
  2M =
  -(1+r_0^2)(1+r_+^2)(1+r_-^2)=
  2r_+^2(1+r_+^2)(1+\alpha^2_+-\epsilon)(1+(1-2\epsilon)r_+^2).
\end{gather}
Given that the black hole charges are determined by three parameters,
the following relation between $r_+$, $\epsilon$, $\alpha_+$ and
$\alpha_-$ holds
\begin{equation}
  (\alpha^2_+-(1-2\epsilon)r_+^2)^2-\alpha_-^4=
  (1-2\epsilon)(1+r_+^2)(1+(1-2\epsilon)r_+^2),
  \label{eq:hyperboleparams}
\end{equation}
so that, for fixed $r_+$ and $\epsilon$, the set of available
$\alpha_+^2$ and $\alpha_-^2$ spans a hyperbola. The requisite of
positive temperature will truncate the hyperbola well before
$\alpha_+^2<1/r_+^2$, which corresponds to $a_1,a_2<1$. We will assume
that $a_1>a_2$, \textit{i.e.}  $\alpha_-^2$ positive. 

Let us now use $z_0=(r_+^2-r_-^2)/(r_+^2-r_0^2)$ as extremality
parameter. We can see that it is proportional to the temperature by
employing \eqref{eq:bhparameters}
\begin{equation}
  2\pi T_+ =
  \frac{r_+(r_+^2-r_-^2)(r_+^2-r_0^2)}{(r_+^2+a_1^2)(r_+^2+a_2^2)}
  = 
  z_0\frac{(1+(3+2\alpha^2_+-2\epsilon)r_+^2)^2}{
    2r_+^3(1+\alpha_+^2-\epsilon)(1+(1-2\epsilon)r_+^2)},
  \label{eq:temperaturez0}
\end{equation}
and it vanishes with $\epsilon$:
\begin{equation}
  z_0=\frac{r_+^2-r_-^2}{r_+^2-r_0^2}=
  \frac{2\epsilon r_+^2}{1+(3+2\alpha_+^2-2\epsilon)r_+^2},\qquad
  \epsilon=z_0\frac{1+(3+2\alpha_+^2)r_+^2}{2(1+z_0)r_+^2}.
\end{equation}
The parametrization of $T_+$ in terms of $z_0$ is relevant for the
near-extremal limit of the monodromy parameters
\eqref{eq:radialmonodromies}, to be discussed below. Dwelling on this
a little further, consider the parametrization of $T_+$ in terms of $\epsilon$ from
\eqref{eq:temperaturez0}
\begin{equation}
  2\pi T_+ =\frac{\epsilon}{r_+} \frac{1+(3+2\alpha^2_+-2\epsilon)r_+^2}{
    (1+\alpha_+^2-\epsilon)(1+(1-2\epsilon)r_+^2)}.
   \label{eq:temperatureepsilon}
\end{equation}
We find that the temperature is generically proportional to
$\epsilon$, but small temperatures are defined relative to $r_+$. This
means that, for small black holes, it is not sufficient to require
small $\epsilon$, but also that $\epsilon \ll  r_+$.  In addition, we
will take $\alpha_\pm$ to be of order one, following the argument at
the beginning of the Section.

With the parametrization above, the $\theta_\pm$ parameters defined by
\eqref{eq:radialmonodromies} have the asymptotic form:
\begin{equation}
  \theta_\pm =\pm\frac{i\Lambda}{z_0}+\frac{\theta_\star}{2},
  \label{eq:zerotmomenta}
\end{equation}
where $\Lambda$ and $\theta_\star$ will have their expansions in $r_+$
given below. Low temperature ($\epsilon\ll 1$), small $r_+$ expansions
of the relevant quantities are 
\begin{equation}
  \theta_0=\omega+(m_1\alpha_1+m_2\alpha_2)r_+
  -3(1+\alpha_+^2-\epsilon)\omega r_+^2+\ldots
  \label{eq:asymptheta0}
\end{equation}
\begin{multline}
  \Lambda =
  -(2(1+\alpha_2^2)-\epsilon(1-\alpha_2^2))\frac{m_1\alpha_1}{2}r_+^2
  -(2(1+\alpha_1^2)-\epsilon(1-\alpha_1^2))\frac{m_2\alpha_2}{2}r_+^2\\
  +(2(1+\alpha_+^2)-(3+\alpha_+^2)\epsilon)\omega r_+^3 +\ldots 
  \label{eq:asymplambda}
\end{multline}
\begin{multline}
  \theta_\star=
  -i(2(1-\alpha_2^2)-\epsilon(1-3\alpha_2^2))\frac{m_1\alpha_1}{4}
  -i(2(1-\alpha_1^2)-\epsilon(1-3\alpha_1^2))\frac{m_2\alpha_2}{4}\\
  +i(2+\epsilon)(1+\alpha_+^2)\frac{\omega r_+}{2} +\ldots
  \label{eq:asympthetas}
\end{multline}
where $\alpha_i=a_i/r_+$, and the ellipsis stands for terms
of higher order in $r_+$ and $\epsilon$, which will not be relevant
for the following analysis. Note that the leading behavior of 
$\Lambda$ and $\theta_\star$
with $r_+$ depends on whether $m_1$ and $m_2$
vanish or not. In particular, if $m_1=m_2=0$, then $\Lambda$ vanishes
as $r_+^3$ and $\theta_\star$ vanishes as $r_+$.
Otherwise, they vanish as
$r_+^2$ and have finite limits, respectively. This behavior does not
depend on whether $\epsilon$, which is proportional to the
temperature, is zero or not. 

The quantization conditions \eqref{eq:monoconditions} can be applied,
along with \eqref{eq:radialconditions} to find the QNMs'
frequencies. For generic temperatures, the study was performed in 
\cite{Barragan-Amado:2018pxh}. The low-temperature limit is a
confluence limit of the Heun equation, given that both $\theta_+$ and
$\theta_-$ diverge as $z_0\rightarrow 0$. The tau function defined in
\eqref{eq:fredholmexpansion} will have to be studied in this limit.

\subsubsection{The confluent limit of the tau function}
\label{sec:confluence}

The Plemelj operator $\mathsf{D}$ which composes the
Fredholm determinant \eqref{eq:fredholmexpansion} has a fairly
straightforward confluence limit. We will assume that the $\sigma$ and
$\kappa$ parameters, defined in \eqref{eq:tildes}, have finite limits
as $z_0\rightarrow 0$, which will be verified \textit{a
  posteriori}. In this case, the hypergeometric functions entering the
parametrices $\mathsf{D}$ can be expanded in an asymptotic series for
small $z_0$ yielding confluent hypergeometric functions 
\begin{equation}
  \Psi(-\sigma,\theta_+,\theta_-;z_0/z) =
  \Psi^{(0)}_c(-\sigma,\theta_\star;i\Lambda/z)
  +z_0\Psi^{(1)}_c(-\sigma,\psi_+,\psi_-;i\Lambda/z)+\ldots
  \label{eq:parametricesc}
\end{equation}
where the confluent parametrix is, defining $\theta_\star=\psi_++\psi_-$:
\begin{gather}
  \Psi^{(0)}_c(-\sigma,\theta_\star;i\Lambda/z)=
  \begin{pmatrix}
    \phi_c(-\sigma,\theta_\star;i\Lambda/z) &
    \chi_c(-\sigma,\theta_\star; i\Lambda/z) \\
    \chi_c(\sigma,\theta_\star; i\Lambda/z) &
    \phi_c(\sigma,\theta_\star; i\Lambda/z)
  \end{pmatrix},
  \nonumber \\
  \phi_c(\pm\sigma,\theta_\star;i\Lambda/z) = 
  {_1F_1}(\tfrac{-\theta_\star\pm\sigma}{2};\pm\sigma;-i\Lambda/z), \\
  \chi_c(\theta_\star,\pm\sigma;-i\Lambda/z) =
  \pm\frac{-\theta_\star\pm\sigma}{2\sigma(1\pm\sigma)}\,
  \frac{i\Lambda}{z}
  \,{_1F_1}(1+\tfrac{-\theta_\star\pm\sigma}{2},2\pm\sigma;-i\Lambda/z).
\end{gather}
It follows that the operator $\mathsf{D}$ is analytic in $z_0$ and
$\Lambda$ 
\begin{equation}
  \mathsf{D}(z_0,i\Lambda)=\mathsf{D}_0(i\Lambda)+z_0 \mathsf{D}_1(i\Lambda)+
  z_0^2\mathsf{D}_2(i\Lambda)+\ldots,
  \label{eq:thisisdeenaught}
\end{equation}
with each term computed recursively in terms of confluent
hypergeometric functions. The first term $\mathsf{D}_0(i\Lambda)$
defines the Fredholm determinant expression for the Painlevé V
transcendent defined in \cite{Lisovyy:2018mnj}. As we will see, we
will not need the corrections $\mathsf{D}_1,\mathsf{D}_2,\ldots$ for
our analysis. 

Let us now study the behavior of the $s$ parameter appearing in
\eqref{eq:fredholmexpansion} as $z_0\rightarrow 0$. The expansion of
the Painlevé VI tau function has the schematic structure
\eqref{eq:taufunctionexpansion} 
\begin{equation}
  \tau = z_0^{A_0}(1-z_0)^{A_1}
  (1 + A_{1,-1}Y ^{-1}z_0+A_{1,0}z_0+A_{1,1}Y z_0+{\cal O}(z_0^2)),
\end{equation}
where $A_0,A_1,A_{1,0},A_{1,\pm 1}$ are monodromy-dependent
coefficients and $Y = \kappa z_0^\sigma$, with $\kappa$ defined in
\eqref{eq:tildes}. The series, save from the $z_0^{A_0}$ term, is
analytic in $z_0$ and meromorphic in $Y$. Finding the non-trivial zero
for $Y$ -- the first condition in \eqref{eq:monoconditions} -- can
then be achieved perturbatively (assuming $\Re \sigma>0$) 
\begin{equation}
  Y=\kappa z_0^{\sigma} = -A_{1,-1}z_0\chi_{VI}(z_0),
  \label{eq:chi6expansion}
\end{equation}
where $\chi_{VI}(z_0)$ is given as a series in $z_0$. The left hand
side can be written as 
\begin{equation}
  \kappa z_0^\sigma=s\Pi z_0^{\sigma}
  \frac{\Gamma(1+\tfrac{i}{2}\Lambda z_0^{-1}+\frac{\sigma}{2})}{
   \Gamma(1+\tfrac{i}{2}\Lambda z_0^{-1}-\frac{\sigma}{2})}
\end{equation}
with $\Pi$ defined as the confluent limit of the product of Gamma
functions in \eqref{eq:tildes}
\begin{equation}
  \Pi=\frac{\Gamma^2(1-\sigma)}{\Gamma^2(1+\sigma)}
  \frac{\Gamma(1+\tfrac{1}{2}(\theta_\star+\sigma))}{
    \Gamma(1+\tfrac{1}{2}(\theta_\star-\sigma))} 
    \frac{\Gamma(1+\tfrac{1}{2}(\theta_1-\theta_\infty+\sigma))
    \Gamma(1+\tfrac{1}{2}(\theta_1+\theta_\infty+\sigma))}{
    \Gamma(1+\tfrac{1}{2}(\theta_1-\theta_\infty-\sigma))
    \Gamma(1+\tfrac{1}{2}(\theta_1+\theta_\infty-\sigma))}.
  \label{eq:thisispi}
\end{equation}
The right hand-side of \eqref{eq:chi6expansion} can now be expanded
\begin{equation}
  A_{1,-1}z_0\chi_{VI}(z_0)= A^{V}_{1,-1} z_0
  \left(\frac{\sigma}{2}+\frac{i\Lambda}{z_0}\right) 
  \left(\chi_{V}(i\Lambda)+{\cal O}(\Lambda^n z^m_0) \right)
  \label{eq:chi0exp}
\end{equation}
where $\chi_{V}(i\Lambda)$ is the series analogous
to \eqref{eq:chi6expansion} that we find from the Painlevé V tau
function. The correction terms to the series are, as argued above, of
the form $\Lambda^nz_0^m$ and hence subleading. Using the property of
the gamma function in the left hand side to cancel the
$\sigma/2+i\Lambda/z_0$ term, we have the expansion
\begin{equation}
  \frac{\Gamma(\tfrac{i}{2}\Lambda z_0^{-1}+\frac{\sigma}{2})}{
    \Gamma(1+\tfrac{i}{2}\Lambda z_0^{-1}-\frac{\sigma}{2})}
  = \left(\frac{i\Lambda}{z_0}\right)^{\sigma-1}\left(
    1+{\cal O}\left(\frac{z_0^2}{\Lambda^2}\right)\right),
\end{equation}
for the argument of the Gamma function not close to the negative
real axis -- which should be expected for real positive
frequencies. Combining the expansions, we find that in the confluent
limit, \eqref{eq:chi6expansion} is replaced by
\begin{equation}
  \kappa z^{\sigma} =s\Pi (i\Lambda)^{\sigma}
  \left(1+{\cal O}(z_0^2\Lambda^{-2})\right).
\end{equation}
In conclusion, we have that in order to compute the first non-trivial
$z_0$ correction, it suffices to consider the Painlevé V
expansion. Corrections to the monodromy parameter are of order
$z^2_0$, and hence less relevant for the confluent limit.

Let us now review the type of correction we have for the tau function
away from the confluence point, where it coincides with the Painlevé V
transcendent. The relevant expansions in $\Lambda$ and $z_0$ are:
\begin{enumerate}
  \item Expansion of the $s$ parameter, which is given in terms
    of powers of $-iz_0/\Lambda$.
  \item Expansion of the parametrices $\Psi_c^{(i)}$
    \eqref{eq:parametricesc}, which are given 
    in terms of hypergeometric functions with monomials in
    $z_0^n(i\Lambda)^m$.
\end{enumerate}
It is not difficult to see that, in the small $\Lambda$ limit, the
first type of expansion will dominate over the second. Disregarding
the contributions of the second type, we have exactly the same type of
expansion as the Painlevé V. This means that the first non-trivial
correction to the accessory parameter problem of the confluent Heun
equation comes in the monodromy parameter, not in the tau function
expansion. In short, the first non-trivial correction is
obtained by expanding the $\kappa$ parameter. 

The final step needed before the calculation can be performed is
solving for the quantization condition for the radial equation,
expressed in terms of the monodromy parameters in
\eqref{eq:radialconditions}. It approaches, as $z_0\rightarrow 0$, 
\begin{equation}
  s=e^{\mp i\pi\sigma}\frac{\sin\frac{\pi}{2}(\theta_\star+\sigma)}{
   \sin\frac{\pi}{2}(\theta_\star-\sigma)}
  \frac{\sin\frac{\pi}{2}(\theta_1+\theta_\infty+\sigma)
    \sin\frac{\pi}{2}(\theta_1-\theta_\infty+\sigma)}{
    \sin\frac{\pi}{2}(\theta_1+\theta_\infty-\sigma)
    \sin\frac{\pi}{2}(\theta_1-\theta_\infty-\sigma)}+\ldots,
  \label{eq:quantizationcondition}
\end{equation}
for $\Re\Lambda>0$ or $\Re\Lambda<0$, respectively. In this case the
convergence is exponential, up to terms of order
$e^{-2\pi|\Lambda|/z_0}$. Again, corrections are (exponentially)
suppressed in terms of $|\Lambda|$. 

\subsubsection{Quasi-normal modes}

With these premises, the solution for the QNMs
follows from the radial quantization condition, expressed in terms of
$s$ in \eqref{eq:quantizationcondition}, and the analysis is actually closely
related to that of \cite{Barragan-Amado:2018pxh}. In the
$z_0\rightarrow 0$ limit, the conditions to solve become  
\begin{gather}
   \label{eq:zeroconfluent}
  \tau_V(\theta_\star,\theta_1,\theta_\infty;\sigma,s;i\Lambda)=0,\\
  c_0=i\Lambda\frac{d}{du}\log\hat{\tau}(\theta_\star-1,\theta_1,\theta_\infty+1;
  \sigma-1,s;i\Lambda)+\frac{1}{4}((\sigma-1)^2-(\theta_\star-1)^2)
  \label{eq:accessoryconfluent}
\end{gather}
with
$\hat{\tau}=\det(\mathbbold{1}-\mathsf{A}\Phi\mathsf{D}_0\Phi^{-1})$
and $\tau_{V}(i\Lambda)=\tilde{C}
(i\Lambda)^{-\frac{1}{4}(\sigma^2-\frac{1}{2}\theta_\star^2)}
e^{-\frac{i}{2}\Lambda\theta_1}\hat{\tau}$ is the Painlevé V tau
function. The Plemelj operator  $\mathsf{A}$ is defined in Appendix
\ref{sec:painleve} and $\mathsf{D}_0$ is in 
\eqref{eq:thisisdeenaught}. The accessory parameter $c_0$ is the limit as
$z_0\rightarrow 0$ of the radial accessory parameter $z_0K_0$, given
by \eqref{eq:accessorykradial} in our application. 

The procedure is now to tackle the condition in
\eqref{eq:zeroconfluent}, which can be
inverted for the zero $\tau_V=0$ in order to solve for $\kappa
z_0^\sigma$ as a series in $\Lambda$. The result is
\begin{subequations}
\begin{equation}
  s\Pi(i\Lambda)^\sigma
  \left(1+{\cal O}(z_0^2)+\ldots\right)
  =i\frac{(\sigma+\theta_\star)((\sigma+\theta_1)^2-\theta_\infty^2)}{
    8\sigma^2(\sigma-1)^2}\Lambda\,\chi(i\Lambda)
  \label{eq:ex}
\end{equation}
where $\chi(i\Lambda)$ is given as a series in $\Lambda$
\begin{equation}
  \chi(i\Lambda)=1+\chi_1(i\Lambda)+\chi_2(i\Lambda)^2+...,
\end{equation}
with
\begin{equation}
    \chi_1=(\sigma-1)\frac{\theta_\star
      (\theta_1^2-\theta_\infty^2)}{\sigma^2(\sigma-2)^2},
\end{equation}
and
\begin{multline}
    \chi_2=\frac{\theta_\star^2(\theta_1^2-\theta_{\infty}^2)^2}{64}
    \left(\frac{5}{\sigma^4}-\frac{1}{(\sigma-2)^4}
    -\frac{2}{(\sigma-2)^2}+\frac{2}{\sigma(\sigma-2)}\right)\\
  -\frac{(\theta_1^2-\theta_{\infty}^2)^2+2\theta_\star^2
    (\theta_1^2+\theta_{\infty}^2)}{64}\left(
    \frac{1}{\sigma^2}-\frac{1}{(\sigma-2)^2}\right)
  \\
  +\frac{(1-\theta_\star^2)(\theta_1^2-(\theta_{\infty}-1)^2)(\theta_1^2
    -(\theta_{\infty}+1)^2)}{128}\left(\frac{1}{(\sigma+1)^2}-
    \frac{1}{(\sigma-3)^2}\right).
\end{multline}
\end{subequations}
The higher order terms can be computed recursively. For our
application, we use the quantization condition
\eqref{eq:quantizationcondition} and, after using $\sin(\pi z)
\Gamma(z)\Gamma(1-z)=\pi$ a number of times, \eqref{eq:ex} now reads
\begin{equation}
  \chi(i\Lambda)=  e^{-\frac{i}{2}\pi\sigma}
  \Theta \Lambda^{\sigma-1}
  \left(1+{\cal O}(z_0^2)+\ldots\right)
  \label{eq:chicondition}
\end{equation}  
where
\begin{equation}
 \Theta =i\frac{\Gamma^2(2-\sigma)\Gamma(\tfrac{1}{2}(\sigma-\theta_\star))
    \Gamma(\tfrac{1}{2}(\sigma-\theta_1+\theta_\infty))
    \Gamma(\tfrac{1}{2}(\sigma-\theta_1-\theta_\infty))}{
    \Gamma^2(\sigma) \Gamma(\tfrac{1}{2}(2-\sigma-\theta_\star))
    \Gamma(\tfrac{1}{2}(2-\sigma-\theta_1+\theta_\infty))
    \Gamma(\tfrac{1}{2}(2-\sigma-\theta_1-\theta_\infty))}.
\end{equation}

The treatment of the second condition \eqref{eq:accessoryconfluent} is
analogous, substituting the value of $\chi(i\Lambda)$ into the
logarithmic derivative and solving for $c_0$ as a function of
$\Lambda$. The result, assuming $\Re\sigma>0$, is
\begin{subequations}
  \label{eq:accessoryc}
\begin{equation}
  c_0=k_0+k_1(i\Lambda)+k_2(i\Lambda)^2+\ldots
  +k_n(i\Lambda)^n+\ldots,
  \label{eq:accessoryc0}
\end{equation}
with the first three terms in the expansion given by
  \begin{equation}  
    k_0=\frac{(\sigma-1)^2-(\theta_\star-1)^2}{4},\quad\quad
    k_1=\frac{\theta_1-1}{2}+\frac{\theta_\star}{2}+
    \frac{\theta_\star(\theta_1^2-\theta_\infty^2)}{4\sigma(\sigma-2)},
    \label{eq:accessoryc01}
  \end{equation}
  \begin{multline}
    k_2=\frac{1}{32}+\frac{\theta_\star^2(\theta_1^2-\theta_{\infty}^2)^2}{64}
  \left(\frac{1}{\sigma^3}-\frac{1}{(\sigma-2)^3}\right)
      +\frac{(1-\theta_\star^2)(\theta_\infty^2-\theta_{1}^2)^2+2\theta_\star^2
        (\theta_\infty^2+\theta_{1}^2)}{32\sigma(\sigma-2)}
      \\-
        \frac{(1-\theta_\star^2)((\theta_\infty-1)^2-\theta_{1}^2)((\theta_\infty+1)^2-
          \theta_{1}^2)}{32(\sigma+1)(\sigma-3)}.
     \label{eq:accessoryc2}
  \end{multline}
\end{subequations}

Each term in the expansion of $\chi$ and $c_0$ is a meromorphic
function of $\sigma$, symmetric under $\sigma\rightarrow
2-\sigma$. There are single poles at $\sigma=3,4,5,\ldots$, with the
pole of order $\sigma=n$ only seen at $k_{n-1}$ term, and a pole
of higher order near $\sigma= 2$, which is present at all orders save
for $k_0$. The asymptotics of the $\chi_n$ \eqref{eq:ex} and $k_n$
\eqref{eq:accessoryc} terms as $\sigma\simeq 2$, reads:
\begin{equation}
  (\sigma-2)\chi_n \simeq k_n=-\mathbf {C}_{n-1}
  \frac{\theta_\star^n(\theta_1^2-\theta_\infty^2)^n}{
    8^n(\sigma-2)^{2n-1}}+\ldots,\quad\quad
  n\ge 2,
  \label{eq:asympmonos}
\end{equation}
where the terms left out are of lower order in $\sigma-2$. The
relation $k_n=(\sigma-2)\chi_n$ no longer holds as one considers less
divergent terms $(\sigma-2)^{-m}$, $m<2n-1$. Finally, $\mathbf{C}_n$
is the $n$-th Catalan number: 
\begin{equation}
  \mathbf{C}_n=\frac{1}{n+1}\begin{pmatrix} 2n \\ n \end{pmatrix}
  = 1, 1, 2,5,14,\ldots
\end{equation}
This remark will be useful when we discuss the CFT description and the
fundamental QNM in the following.

\subsection{QNMs and CFT description of perturbations}

Before delving into the QNMs, let us discuss the
conformal aspects of the perturbations, both in the four-dimensional
picture, and the ``two-dimensional'' point of view encoded in the
preceding analysis. If we take the massless scalar field (at
$\Delta=4$) to represent the scalar sector of gravitational perturbations in
AdS, then the gravitational QNMs can be understood as
resonances created by perturbations of the CFT state represented by the
stress-energy tensor \eqref{eq:holostressenergy}. The holographic
view of the scattering process has been considered in a great number
of papers over the years, with the ideas listed here discussed in
\cite{Hijano:2015zsa} and \cite{daCunha:2016crm}. The starting point
is that scattering amplitudes are encoded in the 4-point function
\begin{equation}
  \langle {\cal V}_{\mathrm{Heavy}}(\infty){\cal
    V}_{\mathrm{Light}}({\bf x}_1){\cal V}_{\mathrm{Light}}({\bf
    x}_2){\cal V}_{\mathrm{Heavy}}(0)\rangle
  \label{eq:4dconformalblock}
\end{equation}
with ${\cal V}_{\mathrm{Heavy}}$ vertex operators associated with the
background charges -- the mass and angular momenta of the black hole
in our case -- and ${\cal V}_{\mathrm{Light}}$ associated with the
local insertion of the scalar perturbations. In a Euclidean setting
the positions of the heavy operators are unambiguous, but in our
Lorentz setting we will take $0$ and $\infty$ to denote past and
future infinity respectively. The correlation functions such as in
\eqref{eq:4dconformalblock} have a rich history in the early
development of CFTs, notably the seminal work of
\cite{Ferrara:1971vh,Ferrara:1972xq,Ferrara:1973vz,Polyakov:1974gs}. The
``partial wave'' decomposition studied there can be thought of as
a higher dimensional conformal block, in which the quantum numbers of
the intermediate states mirror the decomposition of the (scalar) wave
equation in terms of $\mathrm{SO}(4,2)$ quantum numbers associated to the
generators \eqref{eq:so42charges}. 

For pure AdS, it has been known for some time that these higher
dimensional conformal blocks can be factorized in terms of
``two-dimensional'' classical conformal blocks (hypergeometric
functions) \cite{Dolan:2000ut,Dolan:2003hv,Dolan:2011dv}. Now, we want
to view the conditions \eqref{eq:monoconditions} relating the parameters
of the differential equation to monodromy data as a deformed version
of this factorization, valid for the case with non-zero background charges
\eqref{eq:so42charges}. Given that the tau function expansion is given
in terms of conformal blocks \cite{Gamayun:2012ma}, there are indeed
some analogies, although in this case the two-dimensional CFT is
chiral and has unit central charge $c=1$. Furthermore, $z_0$ and
$u_0$ are independent, and in principle complex, variables. 

The latter point may be a consequence of Liouville exponentiation
\cite{Iorgov:2014vla}, and indeed there is a semiclassical
view also relating the accessory parameter of the Heun equation to the
Painlevé transcendent \cite{Litvinov:2013sxa}. In the latter view, the
accessory parameter also appears as the logarithmic derivative of the
two-dimensional conformal block:
\begin{equation}
  \langle
  V_{\Delta_1}(\infty)V_{\Delta_2}(1)\Pi_{\sigma}V_{\Delta_3}(t)V_{\Delta_4}(0)\rangle
  \label{eq:2dconformalblock}
\end{equation}
where $V_{\Delta_i}(z_i)$ denotes each vertex operator, and
$\Pi_{\sigma}$ the projection onto the Verma module constructed on the
primary state, with Liouville momentum parametrized by $\sigma$. The
conformal dimension $\Delta_i$ of each vertex operator 
$V_{\Delta_i}(z)$ is given in terms of the Liouville momentum
$P_i$ and the Liouville coupling constant $b$ as
\begin{equation}
  \Delta_i = \frac{c-1}{24}+P_i^2, \qquad c=1+6Q^2=1+6(b+b^{-1})^2. 
\end{equation}

In this picture, all operators appearing in \eqref{eq:monoconditions}
are ``heavy'' in the sense that they have large Liouville momenta:
\begin{equation}
  P_0=\frac{\theta_-}{b},\qquad P_t=\frac{\theta_+}{b},\qquad
  P_1=\frac{2-\Delta}{b},\qquad P_\infty=\frac{\theta_0}{b},
\end{equation}
which not only provide the Liouville interpretation for the single
monodromy parameters $\theta_i$, but also hint at a deeper connection. If the
black hole absorbs a quantum of energy and angular momenta given by
$\omega$, $m_1$ and $m_2$, respectively, then the increase in its
entropy is given by
\begin{equation}
  \delta S = \frac{\omega-m_1\Omega_{1,+}-m_2\Omega_{2,+}}{b^2T_+}=
  2\pi\frac{\theta_+}{b^2},
\end{equation}
where we recovered Liouville theory's $\hbar=b^2$. Thus, the entropy
gained in the process is, up to the Liouville coupling, 
given by the Liouville momentum. This gives support to the
interpretation of $V_{\Delta_i}$, at least for the inner and outer
horizon, as associated to a \textit{thermal state} in the underlying dual
theory. By assuming the latter to be unitary and modular invariant, we
can use Cardy's formula \cite{Cardy:1986ie}
\begin{equation}
  \frac{S}{2\pi}\simeq \sqrt{\frac{c}{6}\left(L_0-\frac{c}{24}\right)}
\end{equation}
for the entropy of such CFT at conformal dimension
$L_0=\Delta_i$. Assuming the classical limit $b\rightarrow 0$,
and solving for $L_0=\Delta_i$, we have again the interpretation of
$\delta S_i/2\pi b$ as the Liouville momentum. 

The construction outlined above is suitable for the generic case
considered in \cite{Barragan-Amado:2018pxh}, but it works in a
completely analogous way for the confluence limit we considered in
Section \ref{sec:confluence}. In the latter case, the two primary vertex
operators associated to the inner and outer horizons merge into a
Whittaker operator \cite{Gaiotto:2009ma}. The associated conformal
block is then an irregular one \cite{Nagoya:2015cja}, instead of
\eqref{eq:2dconformalblock}. The argument presented here about the
nature of the intermediate states does have a direct parallel, though,
and for further details we refer to the discussion in 
\cite{CarneirodaCunha:2019tia}. 

\subsection{Fundamental QNM}
\label{sec:fundamentalmode}

The calculation of the fundamental mode is technically very similar
to the corresponding one in \cite{Barragan-Amado:2018pxh}, where one
solves for the first eigenfrequency $\omega_{1,0,0,0}$ by applying the
conditions \eqref{eq:monoconditions} given the quantization of the
angular and radial monodromy parameters \eqref{eq:sigmaell} and
\eqref{eq:sigman}. From the discussion above, the corresponding task
at low temperature involves solving the equations
\eqref{eq:chicondition} and \eqref{eq:accessoryc}.

\begin{center}
\begin{figure}[tb]
  \mbox{\includegraphics[width=\textwidth]{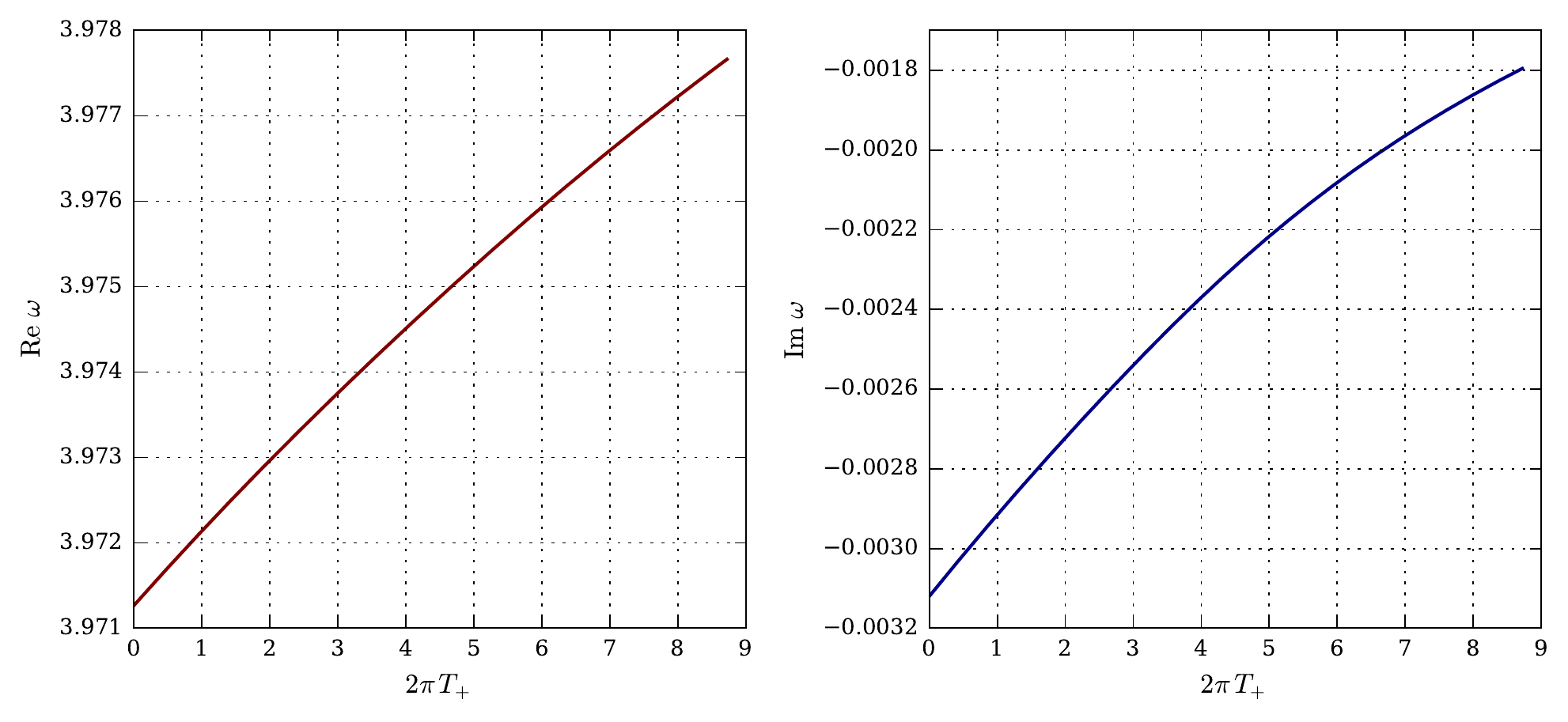}}
  \caption{The fundamental QNM frequency
    $\omega_{1,0,0,0}$ as a function of the outer horizon temperature
    $T_+$ for a Kerr-AdS$_5$ black hole with fixed $r_+=0.03$ and
    $\alpha_{+}^2=\sqrt{2}$. We note that the temperature has the
    effect of increasing both the real and imaginary parts of the
    frequency, up to the maximal value $2\pi (T_+)_{\mathrm{max}}
    \simeq 8.7$, corresponding to $r_-=0$, or $\epsilon=1/2$ in
    \eqref{eq:temperatureepsilon}.} 
  \label{fig:fundamentalmode}
\end{figure}
\end{center}

As suggested by Fig. \ref{fig:fundamentalmode}, for generic values of
$M$, $a_1$ and $a_2$ a smooth zero-temperature limit for the
fundamental QNM frequency exists. At small $r_+$, we can
work out asymptotic formulas, with 
\begin{equation}
  i\Lambda = i\lambda r_+^3,\qquad
  \theta_\star = i\phi_\star r_+,\qquad
  \sigma =2-\nu r_+^2,
  \label{eq:monoexpansion}
\end{equation}
where the values of $\lambda$ and $\phi_\star$ can be read from
\eqref{eq:asymplambda} and \eqref{eq:asympthetas}, by 
setting $m_1=m_2=0$. We find 
\begin{equation}
  \phi_\star = 
  \frac{1}{2}(2+\epsilon)(1+\alpha_+^2)\Delta +\ldots,
  \qquad
  \lambda = (2(1+\alpha_+^2)-(3+\alpha_+^2)\epsilon)\Delta + \ldots, 
\end{equation}
where we used the fact that $\omega$ is approximately $\Delta$ in the
$r_+\rightarrow 0$ limit. As a matter of fact, we will assume that
\begin{equation}
  \theta_0=\omega -3(1+\alpha_+^2-\epsilon)\omega r_+^2 +\ldots
  = \Delta -\beta r_+^2,
\end{equation}
where we introduced the $\beta$ parameter to measure the difference
between $\Delta$ and the monodromy parameter $\theta_0$. Note also
that the expansion of $\theta_0$ implies that $\beta$ is related to
the correction to the QNM frequency at small $r_+$
\begin{equation}
  \omega = \Delta -(\beta-3(1+\alpha_+^2-\epsilon)\Delta)r_+^2+
  {\cal O}(\epsilon^2,r_+^4).
  \label{eq:firstcorrectomega}
\end{equation}
Finally, we remark that, at low temperature we have $\epsilon\ll
r_+\ll 1$, and \eqref{eq:hyperboleparams} implies that $\alpha_1$ and
$\alpha_2$ are related to $r_+$, 
\begin{equation}
  \alpha_1^2\alpha_2^2=(1-2\epsilon)((1+r_+^2)^2-2r_+^2\alpha_+^2)+{\cal
    O}(\epsilon^2).
  \label{eq:relationalphas}
\end{equation}

We will now consider the conditions \eqref{eq:zeroconfluent} and
\eqref{eq:accessoryconfluent}, encoded in the equations
\eqref{eq:chicondition} and \eqref{eq:accessoryc}, as equations
determining $\beta$ and $\nu$ for small $r_+$. First, the $\Theta$
parameter in \eqref{eq:chicondition} has the small $r_+$ expansion for
$\ell=0$: 
\begin{equation}
  \Theta =
  \frac{\phi_\star}{2r_+^3\nu^2}\frac{(\nu+\beta)}{(\nu-\beta)}
  (\Delta-1)
  \left( 1 +i\frac{\nu}{\phi_*}r_++{\cal O}(\epsilon r_+^2,r_+^2)\right).
  \label{eq:Thetaexpansion}
\end{equation}
As anticipated by the argument in the preceding sections, the
$\epsilon$ corrections are accompanied by a factor of $r_+^2$ and are
always subdominant in the small $r_+$ limit.

The accessory parameter $c_0$ as in \eqref{eq:accessoryc} and the
$\chi$ function defined in \eqref{eq:chicondition} both receive
contributions from all orders in 
$z_0$ in the small $r_+$ limit. This is due to the proximity of
$\sigma$ to the critical value $2$, as further clarified in the next
subsection. Contributions to $c_0$ and $\chi$ can be resummed using
the generating formula for the Catalan numbers
\begin{equation}
  1+x+2x^2+5x^3+14x^4+\ldots = \sum_{n=0}^\infty \mathbf{C}_nx^n=
  \frac{1-\sqrt{1-4x}}{2x}.
\end{equation}

The result for $\chi$ can then be written as
\begin{equation}
  \chi 
  =\frac{1}{2}+\frac{\phi_\star(\theta_0^2-\theta_1^2)}{8\nu^2}\lambda+
  \frac{1}{2}\sqrt{1+\frac{\phi_\star(\theta_0^2-\theta_1^2)}{2\nu^2}
    \lambda} 
  -\frac{(\theta_0^2-\theta_1^2)^2\lambda^2}{64\nu^2
    \sqrt{1+\frac{\phi_\star(\theta_0^2-\theta_1^2)}{2\nu^2}\lambda
    }}r_+^2+
  {\cal O}(r_+^4), 
\end{equation}
where $\theta_1=2-\Delta$. At this order in $r_+$, we can substitute
$\theta_0^2-\theta_1^2=4(\Delta-1)+{\cal O}(r_+^2)$. The first
non-trivial correction in $\epsilon$ can be obtained by just
considering the first correction in the parameters $\lambda$ and
$\phi_\star$. The expansion for the accessory parameter $c_0$ in the
right-hand side of \eqref{eq:accessoryc0} can also
be resummed in the same way: 
\begin{equation}
  c_0=\frac{i}{2}\phi_\star r_++\frac{1}{4}\phi_\star^2r_+^2 -
  \frac{\nu r_+^2}{2}\sqrt{1+
    \frac{\phi_\star(\theta_0^2-\theta_1^2)}{2\nu^2}\lambda}+
  \frac{i}{2}(\theta_1-1)\lambda r_+^3+{\cal O}(r_+^4).
  \label{eq:accessoryswave}
\end{equation}

The same parameter $c_0$ is defined as $\lim_{z_0\rightarrow
  0}z_0K_0$ and can be computed from \eqref{eq:accessorykradial} and
\eqref{eq:lambda_ell} by taking the appropriate $r_-\rightarrow r_+$
limit. In the $\epsilon\ll r_+\ll 1$ limit and for $\ell=0$, this
provides for the left-hand side of equation \eqref{eq:accessoryc0},  
\begin{multline}
c_0 =\frac{i}{2}\phi_\star r_++\frac{1}{4}\phi_\star^2r_+^2
-\frac{1+\alpha_+^2}{2}\Delta(\Delta+2)r_+^2\\
+\frac{1}{2}\Delta(\Delta+2)\epsilon r_+^2+
\frac{i}{2}(\theta_1-1)\lambda r_+^3
+{\cal O}(\epsilon^2,\epsilon r_+^4, r_+^4),
\end{multline}
which, equated to the right-hand side of \eqref{eq:accessoryswave}, leads to an
equation for $\nu$ as in \eqref{eq:monoexpansion} 
\begin{equation}
  \sqrt{\nu^2+2\lambda\phi_\star(\Delta-1)} =
  (1+\alpha_+^2)\Delta(\Delta+2)
  -\Delta(\Delta+2)\epsilon+\ldots,
\end{equation}
where the terms left out scale with $\epsilon r_+^2$ and
$\epsilon^2$.  Taking into account \eqref{eq:relationalphas}, this
equation can be solved to yield, at first non-trivial order in
$\epsilon$ and $r_+$,  
\begin{equation}
  \nu =(1+\alpha_+^2)\Delta
  \sqrt{\Delta^2+8}\left(1-\frac{\Delta^2+2\Delta+6}{
      (1+\alpha_+^2)(\Delta^2+8)}\epsilon\right)
  +{\cal O}(\epsilon^2,r_+^2,\epsilon r_+^2)
  \label{eq:smallnu}
\end{equation}
and likewise for $\chi$
\begin{equation}
  \chi 
  =\frac{(\Delta+2+\sqrt{\Delta^2+8})^2}{4(\Delta^2+8)}
  \left(1-\frac{(\Delta+2)(\sqrt{\Delta^2+8}-\Delta-2)}{
      (1+\alpha_+^2)(\Delta^2+8)}\,
    \epsilon \right)
  +{\cal O}(\epsilon^2,r_+^2,\epsilon r_+^2).
  \label{eq:smallchi}
\end{equation}
Finally, by substituting \eqref{eq:smallnu} and \eqref{eq:smallchi}
into \eqref{eq:chicondition}, we can compute $\beta$
\begin{multline}
  \beta = (1+\alpha_+^2)(\Delta(\Delta+2)+2i\Delta(\Delta-1)r_+)
  -\Delta(\Delta+2)\epsilon\\
  -i(3+\alpha_+^2)\Delta(\Delta-1) \epsilon r_+
  +{\cal O}(r_+^2,r_+^2\log r_+,\epsilon r_+^2,\epsilon^2).
\end{multline}
\begin{center}
  \begin{figure}[tb]
    \includegraphics[width=0.95\textwidth]{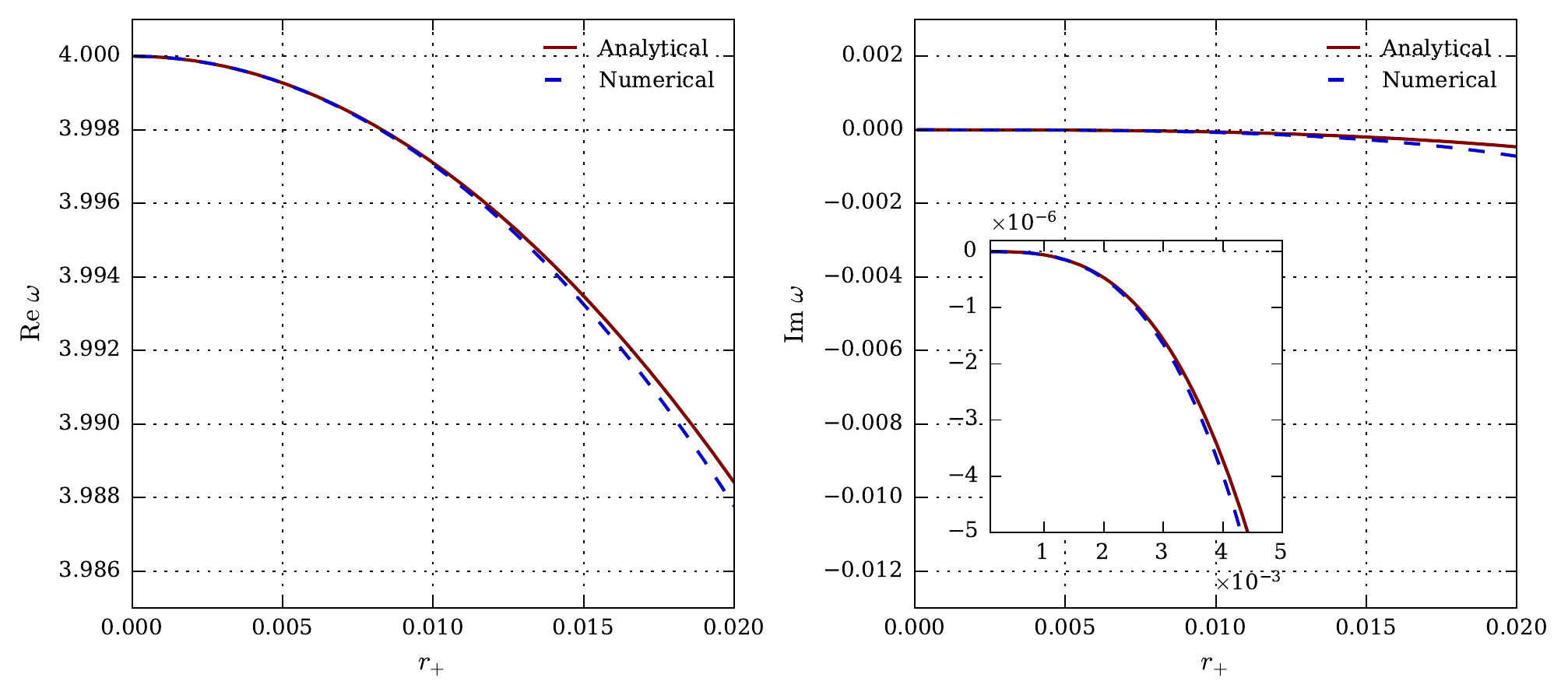}
    \caption{The expression for the fundamental mode
      $\omega_{1,0,0,0}$ given by \eqref{eq:fundomega} compared with
      numerical results for $\epsilon=10^{-6}$ and
      $\alpha_+^2=\sqrt{2}$.}
    \label{fig:fundomega}
  \end{figure}
\end{center}
We remark that now, due to the presence of the $\Lambda^{\sigma-1}$
term in \eqref{eq:chicondition}, there will be non-analytical
corrections of the type $r_+^2\log r_+$ which, however, are
subdominant for small $r_+$. Finally, using the relation between
$\beta$ and $\omega$ given by \eqref{eq:firstcorrectomega}, we find
the first corrections to the fundamental QNM frequency as a function
of $r_+$ and $\epsilon$,
\begin{multline}
  \omega_{1,0,0,0} = \Delta - (1+\alpha_+^2)\Delta(\Delta-1)r_+^2
  -2i(1+\alpha_+^2)\Delta(\Delta-1)r_+^3\\
  +\Delta(\Delta-1)\epsilon r_+^2
  +i(3+\alpha_+^2)\Delta(\Delta-1)\epsilon r_+^3\\
   +{\cal O}(r_+^4,r_+^4\log r_+^2,\epsilon r_+^4,\epsilon^2r_+^2)
   \label{eq:fundomega}
\end{multline}
where the subscript stands for $n,\ell,m_1$ and $m_2$ quantum
numbers. 

The imaginary part of the frequency controls the relaxation times of
the perturbation in the dual CFT. Equation \eqref{eq:fundomega}
is valid for $\epsilon\ll r_+\ll 1$, with $\alpha_+$ constant and of
order ${\cal O}(1)$. In this limit, the effect of non-zero temperature is 
accounted for by the $\epsilon$-dependent terms, which, in agreement
with Fig. \ref{fig:fundamentalmode}, induce an increase of the 
real and imaginary parts of the frequency, as $\epsilon$
increases. Despite having a ``reinforcing'' effect, making
perturbations longer-lived, the temperature does not have a strong
enough influence to induce instabilities. For perturbations above the
unitarity bound $\Delta>1$, the fundamental mode is stable at zero or
small enough temperature. These results complement the
finite-temperature result of \cite{Barragan-Amado:2018pxh}.

\subsubsection{Catalan numbers and intermediate CFT levels}

The two-dimensional version of \eqref{eq:4dconformalblock}, depicting
the perturbation of the black hole background, was studied in
\cite{Fitzpatrick:2015foa}, where it was shown that the vacuum block,
where the internal dimension is zero, was given by the generating 
function for the Catalan numbers in the heavy-light semiclassical
limit, where two of the insertions have high conformal dimensions
$\Delta_i\propto c\rightarrow \infty$, with $c$ being the central charge.

There is indeed a clear parallel to what is being counted in the
fundamental mode calculations in the last subsection. The
finite-temperature version of the calculation outlined in
\cite{Barragan-Amado:2018pxh} does have this hierarchy 
between the vertex operators, since the ones associated with the inner
and outer horizons do become ``light'' in the small black hole
regime, and the residue at $\sigma=2$ does indeed correspond to the
identity operator as intermediate state in
\eqref{eq:4dconformalblock}.

Yet, there are subtleties. We have already remarked that 
\eqref{eq:2dconformalblock} is a chiral conformal block. A
difference from \cite{Fitzpatrick:2015foa} is that in principle all operators in
\eqref{eq:2dconformalblock} are ``heavy'', in the sense that their
conformal dimensions scaling with $c$. The parametric hierarchy
between operators arises from the parameter $r_+$, which can be made
small. It is a combined effect of the position of the insertion
and the scaling of the dimensions of the ``light'' operators. The
corresponding calculation using the Virasoro algebra seems
straightforward, but lies outside the scope of this work. 

Another difference comes from the confluence limit taken as the
temperature goes to zero. In this case the limiting form of the
Liouville momenta for the inner and outer horizon is given by
\eqref{eq:zerotmomenta}, and the conformal dimensions of
the operators associated to $\theta_\pm$ are in fact much larger than
$c$. The ``light'' operator is the resulting operator coming from the
leading term of the OPE for the operators related to the inner and
outer horizons: 
\begin{equation}
  V_{P_3}(t)V_{P_4}(0)\sim
  V_{P}(0)+\ldots
\end{equation}
where we label the operators by the Liouville momentum, and
$P=P_3+P_4$. As the first descendant of the resulting module $[V_{P}]$ one
finds the Whittaker operator \cite{Gaiotto:2009ma,Nagoya:2015cja}
\begin{equation}
  V_{\theta_\star/b,i\Lambda}(0)\sim
  :e^{2\theta_\star/b\phi_{L}+i\Lambda \partial \phi_{L}}:
  \label{eq:whittakerop}
\end{equation}
which is labelled by the confluence parameters. We used the
Feigin-Fuchs parametrization of the Liouville field $\phi_{L}$ (see
\cite{Nagoya:2015cja} or \cite{CarneirodaCunha:2019tia}) in the
right-hand side of \eqref{eq:whittakerop} to compare our notation with
the literature. In this case, the parameters in the regime
investigated here do warrant the ``light'' requisite of
\cite{Fitzpatrick:2015foa} since they are proportional to $r_+$ and 
$r_+^3$, respectively, as can be checked in \eqref{eq:asymplambda} and
\eqref{eq:asympthetas}. In light of this, the results obtained in 
Sec. \ref{sec:fundamentalmode} point to an analogous simplification of
the conformal block in terms of the Catalan number generating function
when the two light operators in the analysis of \cite{Fitzpatrick:2015foa} are
substituted by an irregular light operator. For more on the relation
between irregular conformal blocks and the Painlevé V tau 
function we recommend \cite{Lisovyy:2018mnj}.

\subsection{Higher quasi-normal modes}
\label{sec:higherqnms}

Higher QNMs, with $\ell\geq 1$, allow for non-zero
values of $m_1$ and $m_2$. In turn, these multiply the angular
velocities $\Omega_{i,+}$, which are given in our parametrization by
\begin{equation}
  \Omega_{i,+}=\frac{a_i(1-a_i^2)}{r_+^2+a_i^2}=
  \frac{\alpha_i(1-\alpha_i^2r_+^2)}{r_+(1+\alpha_i^2)},
\end{equation}
with $\alpha_i=a_i/r_+$, $i=1,2$. As explained before, at small $r_+$
the range of allowed $a_i$ is capped by $r_+$, so it makes sense to
consider fixed values of $\alpha_i\lesssim 1$. Then in the small $r_+$
regime, the velocities $\Omega_{i,\pm}\sim 1/r_+$ become very large
and the monodromy parameter \eqref{eq:radialmonodromies}
\begin{equation}
  \theta_+ = \frac{i}{2\pi}\left(
  \frac{\omega-m_1\Omega_{1,+}-m_2\Omega_{2,+}}{T_+}\right)
  \label{eq:thetaplus}
\end{equation}
varies so that its imaginary part can become negative. Note that $r_+$ can
alternatively be thought of as the ratio between the outer horizon and
the AdS radius.

As a matter of  
fact, it has been long established from heuristic arguments that
combinations of $\omega, m_1$ and $m_2$ such that $\Im\theta_+<0$
display stimulated emission -- the Zeldovich-Starobinski effect -- see
\cite{Brito:2015oca} for a detailed account. When $\Im\theta_+$ is
negative, we have the \textit{superradiance window} and unstable modes
are bound to appear. The instability was 
anticipated by the thermodynamical analysis \cite{Hawking:1999dp} and
first observed numerically for four-dimensional black holes in
\cite{Cardoso:2004hs}. The effect is, however, tiny and hard to study 
using conventional numerical methods, even at finite temperature
\cite{Cardoso:2006wa}. Our results in \cite{Barragan-Amado:2018pxh}
provided supporting evidence for the positivity of the imaginary part
of the $\ell=1,m_1=1, m_2=0$ QNM frequency in the small $r_+$
limit, based on the quantization conditions \eqref{eq:sigmaell} and
\eqref{eq:sigman}. 

\begin{center}
  \begin{figure}[tb]
    \includegraphics[width=\textwidth]{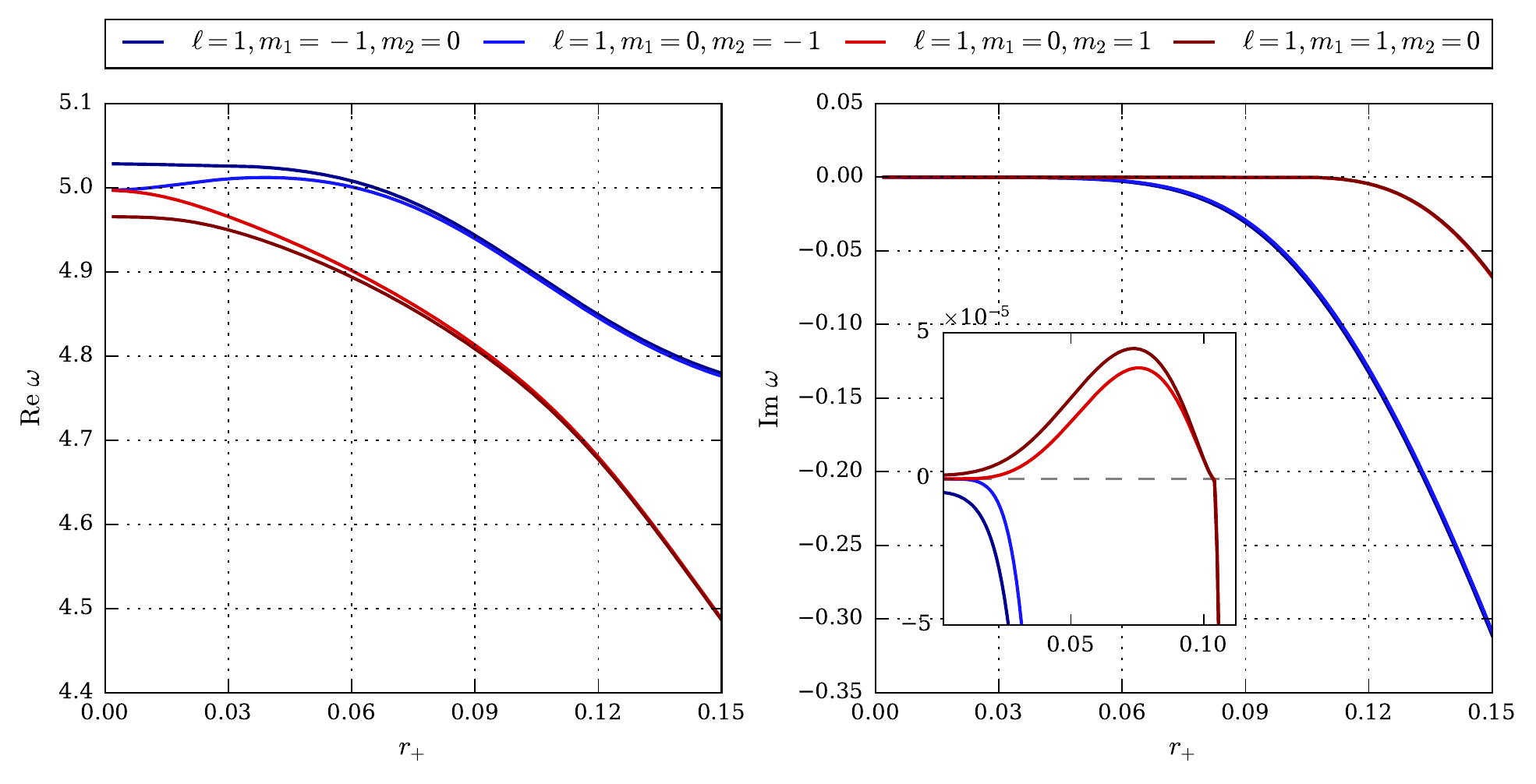}
    \caption{The $\ell=1$ modes for
      $a_{12}=a_1^2-a_2^2=0.001$ as a function of $r_+$ and fixed,
      very small temperature $T_+\sim 10^{-8}$. The modes
      $m_1=1,m_2=0$ (dark red) and $m_1=0,m_2=1$ (light red)
      display a small positive imaginary part for values where
      $\Im\theta_+<0$ (inset).} 
    \label{fig:qnm1110}
  \end{figure}
\end{center}

The low temperature version of the calculation for QNMs in
\cite{Barragan-Amado:2018pxh} follows an analogous
strategy, based on \eqref{eq:chicondition} and
\eqref{eq:accessoryc}. Since the final analytical expression for the 
imaginary part of the $\ell\ge 1$ QNM frequencies is not particularly
illuminating, we will limit ourselves to present 
some numerical results and discuss the qualitative features of the
analyzed modes. The numerical analysis was conducted using an
arbitrary-precision implementation of the Painlevé VI and V tau
functions with a Fourier truncation of the operators $\mathsf{A}$,
$\mathsf{D}$ and $\mathsf{D}_0$ at $N=32$ levels, using the
\href{https://arblib.org}{Arb} library at 
$96$-digit precision\footnote{The
  implementation of the Painlevé VI and V tau functions in
  \href{http://julialang.org}{Julia} programming language can be
  obtained in \href{https://github.com/strings-ufpe/painleve}{
    \texttt{https://github.com/strings-ufpe/painleve}}. The authors
  thank O. Lisovyy for clarification on the details of the truncation
  in a private communication.}. The library is designed to control
numerical uncertainties by performing calculations with intervals of
complex numbers. In all of the analysis the intervals were too small
to be of significance, of order $~10^{-12}$, and they will be omitted
from the analysis.

In Fig. \ref{fig:qnm1110} we display a typical plot of QNMs frequencies
as a function of $r_+$ for $\ell=1$. One sees that modes with
either $m_1$ or $m_2$ positive are unstable for small values of $r_+$
and moderate values of $a_1$ and $a_2$, of which the example shown is
characteristic. From the superradiant condition $\Im\theta_+<0$, one sees
that there are two competing factors determining whether the mode will
be unstable or not: we have the positive contribution from the real
part of the frequency and the negative contribution from the angular
velocity. Now, as one turns on the rotation parameters $a_1$ and
$a_2$, the real part of the frequency gets negative corrections that
overwhelm the growth in angular velocity, effectively reducing the
superradiant window. The ``kink'' in the zoomed in inset as
$\Im\theta_+$ crosses to positive values comes from the change of 
sign of $\Lambda$ in \eqref{eq:zerotmomenta}. In the zero temperature
limit, this change of  sign will induce a discontinuity in the $s$
value obtained from the quantization condition
\eqref{eq:quantizationcondition}. 

\begin{center}
  \begin{figure}[tb]
    \includegraphics[width=\textwidth]{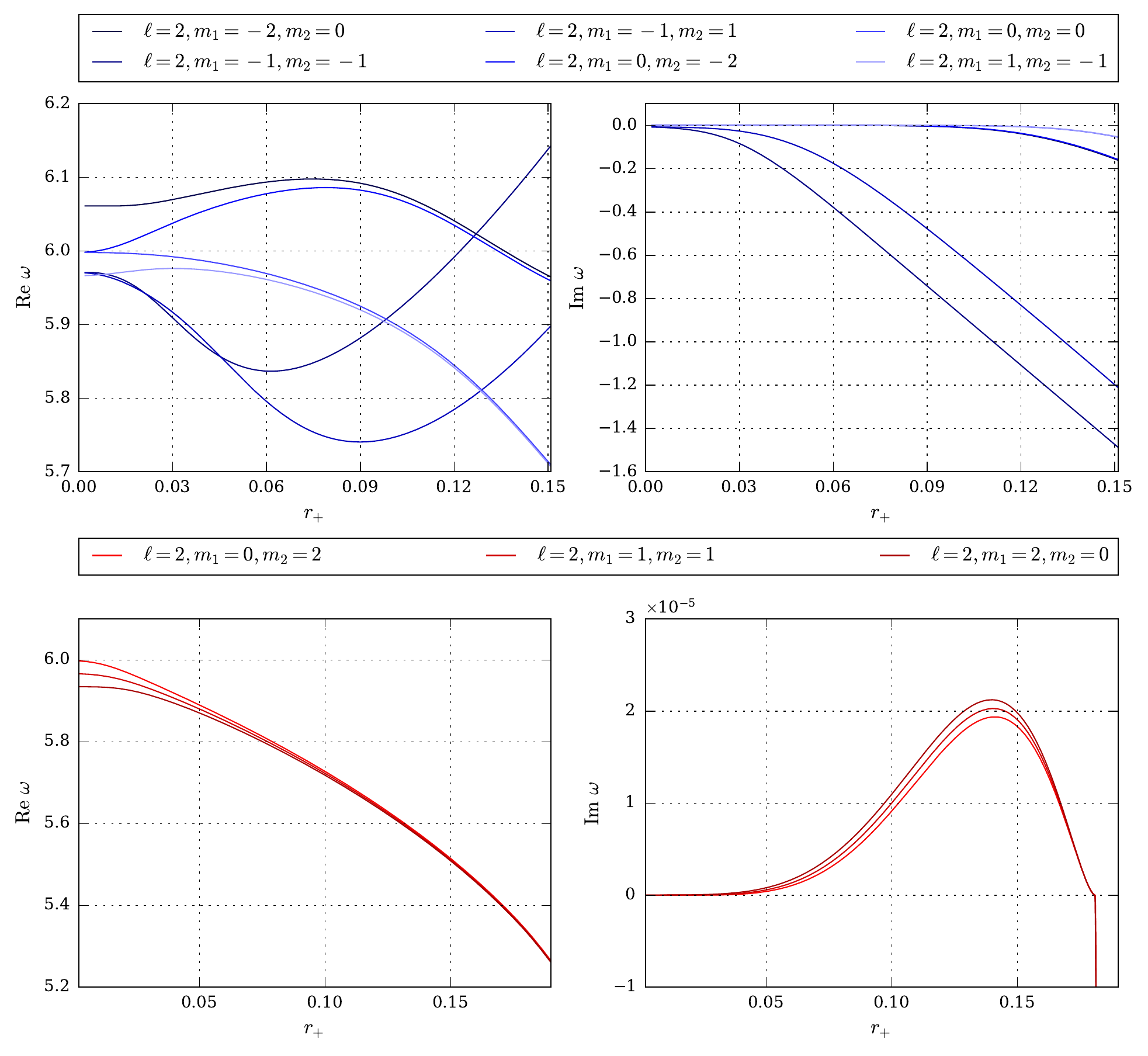}
    \caption{Stable (top, blue) and unstable (bottom, red) modes for
      $\ell=2$ and  $a_1^2-a_2^2=0.001$ as functions of $r_+$. Like in
      the $\ell=1$ case, the positive imaginary part is very small and
      restricted to the condition $\Im\theta_+<0$.}
    \label{fig:qnm2200}
  \end{figure}
\end{center}

From the conformal block point of view, the calculation of the
QNMs is truly perturbative in the CFT levels, with the
internal momentum centered at $\sigma = 2+\ell-\nu_\ell r_+^2$. It is
interesting to note that these should correspond in the
$r_+\rightarrow 0$ limit to degenerate ``heavy'' Liouville
operators at integer $\sigma\simeq 2+\ell$. As a matter of fact, the latter
values for $\sigma$ do indeed correspond to poles in the conformal
block expansion at level $t_0^{\ell+1}$. It would be interesting to investigate
whether this alternative approach could lead to a simplified analytic study of
these modes. 

\begin{center}
  \begin{figure}[tb]
    \includegraphics[width=\textwidth]{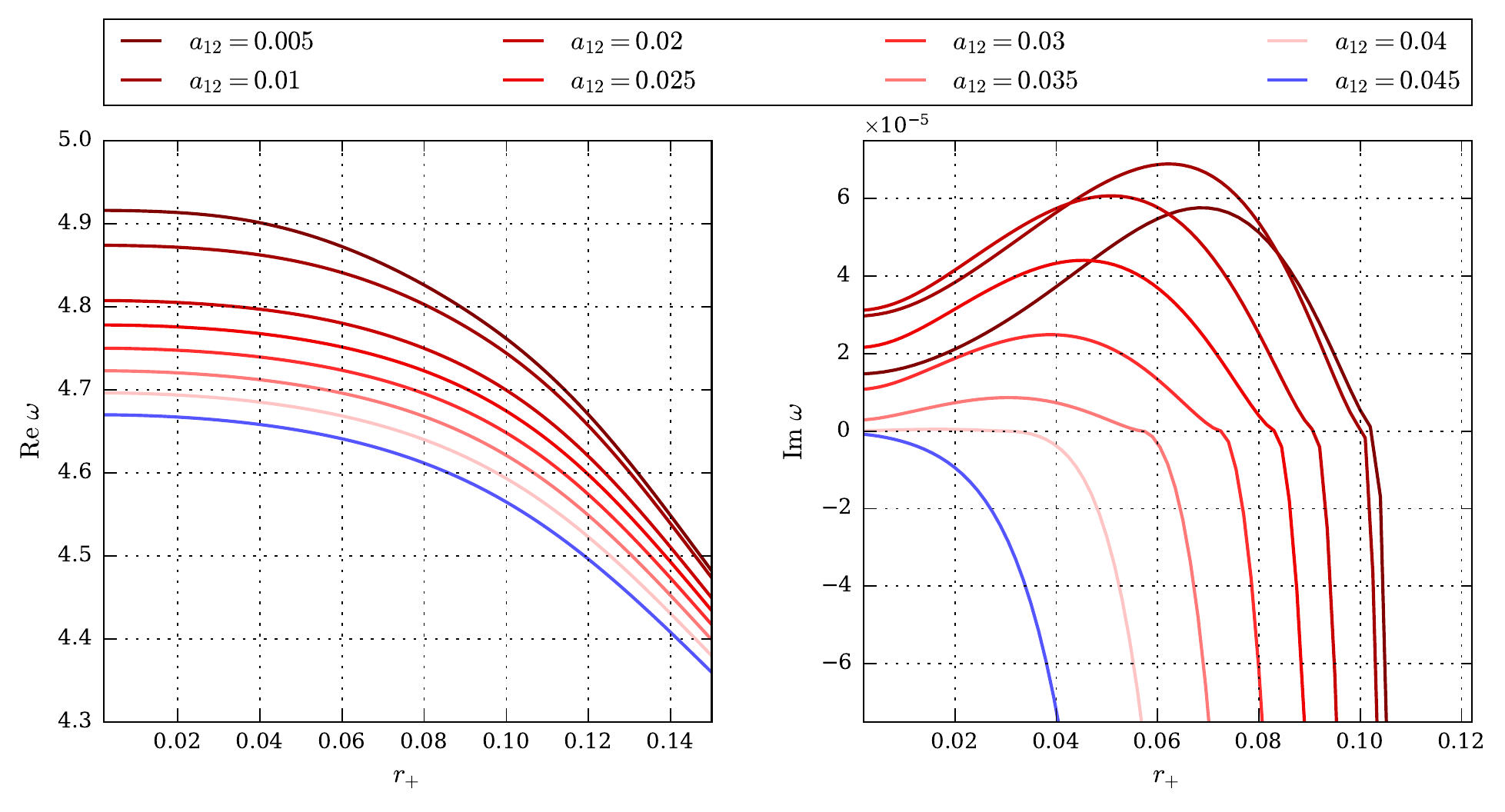}
    \caption{The frequencies of the unstable modes for $\ell=1$ as a
      function of $r_+$ for various values of
      $a_{12}=a_1^2-a_2^2$. Note that the superradiant window induced
      by the condition $\Im\theta_+<0$ becomes smaller with 
      increasing $a_{12}$ due to the larger reduction of the real part
      of the eigenfrequency.}
      \label{fig:qnm1delta}
  \end{figure}
\end{center}

\begin{center}
  \begin{figure}[tb]
    \includegraphics[width=\textwidth]{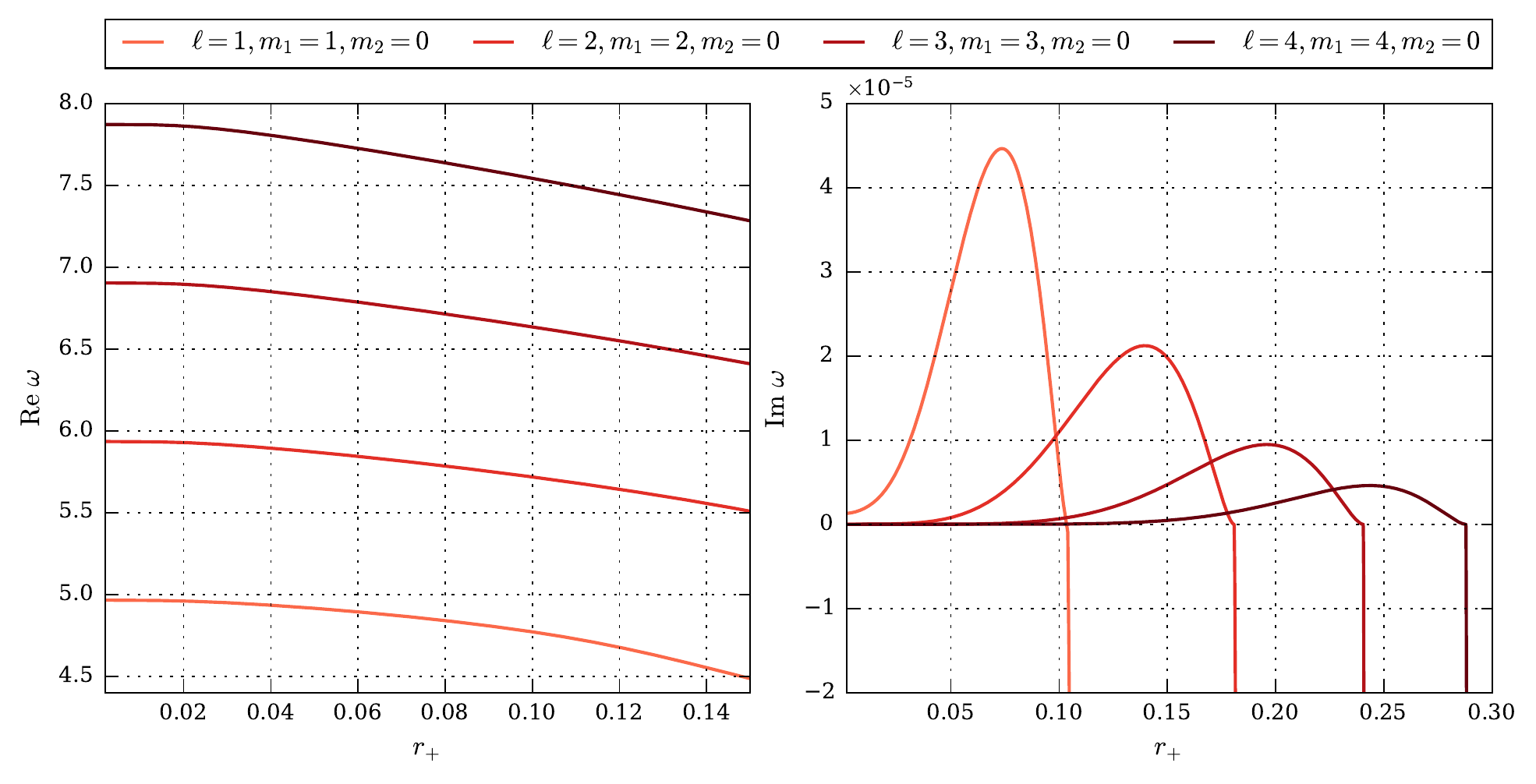}
    \caption{The most unstable mode $m_1=\ell$ as a
      function of $r_+$ for $\ell=1,2,3,4$. Note that the smaller
      decay rate for larger values of $\ell$ is accompanied by a
      larger width of the superradiant window. The black hole is
      expected to decay through different values of $\ell$, $m_1$ and
      $m_2$ depending on the values of its charges.}
       \label{fig:qnm123ell}
  \end{figure}
\end{center}

In Fig. \ref{fig:qnm2200} we display the $\ell=2$ case. Again, the
unstable modes correspond to positive $m_1$ and/or $m_2$, and the
region of instability is determined by the superradiant window
$\Im\theta_+<0$. The comparison to $\ell=1$ -- see also
Fig. \ref{fig:qnm1delta} -- shows that the decay rate of the most
unstable mode \cite{Hawking:1999dp}, \textit{i.e.}, the one with largest
$m_1\Omega_{1,+}+m_2\Omega_{2,+}$ decreases with $\ell$. Indeed, 
Fig. \ref{fig:qnm123ell} shows that the maximum of
the decay rate as a function of $r_+$ does decrease from $\ell=1$ to
$\ell=4$, making those higher $\ell$ states longer-lived. If this
trend continues for larger $\ell$, it would mean that 
the decaying profile of the scalar emissions from the black hole is a
combined effect of the angular dependence of the coupling between the
black hole and the scalar perturbations on the one hand and the half-life of the
perturbation as a function of $\ell$, and the radial
eigenvalue $n$ on the other.  

One should also note from Fig. \ref{fig:qnm1delta} that the
superradiant window appears to decrease as 
$a_{12}=a_1^2-a_2^2$ increases, indicating that, as far as the
contributions to $\theta_+$ are concerned, the negative correction to the
real part of the eigenfrequency in \eqref{eq:thetaplus} dominates over
the effect of the increased angular rotation. So, as $a_{12}$ increases, 
the real part of $\Im\theta_+$ turns positive. In fact, around the
value $a_{12}\simeq 0.045$ there is no superradiance for 
$\ell=1$, as can be inferred from Fig. \ref{fig:qnm1delta}.  This
is in contrast with the fragmentation phenomenon, which is expected at
sufficiently distorted -- \text{i.e}, highly rotating --
five-dimensional black holes as argued in \cite{Emparan:2003sy}.

\section{Holographic decay}
\label{sec:decay}

Given the results of the preceeding sections for the QNMs and their
instability, we may reflect on the fate of the corresponding state 
in the putative dual CFT. The instabilities signal that the state
associated to the black hole will decay. The particular features of the
decay, such as its rate and final products, will of course depend on
the coupling between the black hole and the perturbation fields, which in
holography can be read from the stress-energy tensor. For the
scalar type of perturbations considered here we can deduce an
interaction Hamiltonian of the sort ${\cal H}_{\mathrm{int}}=\lambda
h^{ab}T_{ab}$. As we learned from the discussion above, the imaginary
part of the QNMs frequencies remains fairly constant with $\ell>0$, with
the most unstable mode at given $\ell$ being
the one with the largest $m_1\Omega_{1,+}+m_2\Omega_{2,+}$. 

Let us first consider the case where the field theory lives in the
global boundary $\mathbb{R}\times S^3$. The energy density profile can
be read from \eqref{eq:holostressenergy} 
\begin{equation}
  4\pi G_{N,5} \bar{\rho} = \frac{3M}{2\bar{\Delta}_\theta^2}=
  \frac{3M}{2}\left(1+2(a_1^2\sin^2\bar{\theta}+a_2^2
    \cos^2\bar{\theta})+\ldots\right).
  \label{eq:energylexp}
\end{equation}
We interpret the expansion in the right-hand side to be the
contribution of the higher $S^3$ spherical harmonics. From
\eqref{eq:energylexp} we conclude that the generation of the $\ell$-th
mode will be dampened by a factor of $a_i^\ell$. Since $a_1$ and $a_2$ are
parametrically small, of order $r_+$, the corresponding
five-dimensional spheroidal harmonics can be approximated by their
zero rotation counterparts, \textit{i.e}, the three-dimensional
spherical harmonics \cite{Lindblom:2017maa}  
\begin{multline}
  Y_{\ell}^{m_1,m_2}(\bar{\theta},\bar{\phi}_1,\bar{\phi}_2)=
  \sqrt{\frac{\ell+1}{2\pi^2}}
  \sqrt{\frac{((\ell+m_1+m_2)/2)!((\ell-m_1-m_2)/2)!}{
      ((\ell+m_1-m_2)/2)!((\ell-m_1+m_2)/2)!}}
  \times\\
  (\sin\bar{\theta})^{m_1}(\cos\bar{\theta})^{m_2}
  P_{\frac{1}{2}(\ell-m_1-m_2)}^{(m_1,m_2)}(\cos 2\bar{\theta})
  e^{im_1\bar{\phi}_1+im_2\bar{\phi}_2},
\end{multline}
where $P_{n}^{(a,b)}(z)$ are the Jacobi polynomials. At given $\ell$, the
most unstable mode has the energy dependence proportional to the
$tt$-component of the scalar stress-energy tensor, in turn roughly
proportional to the absolute value squared of the field itself 
\begin{equation}
  \bar{\rho}_{\mathrm{fluc.}}\propto\left|\Phi_{\ell,m_1=\ell,m_2=0}\right|^2 =
  \left|e^{-i\omega(\ell)\bar{t}}Y_{\ell}^{\ell,0}\right|^2
  \propto e^{2\Im\omega(\ell) \bar{t}}\sin^{2\ell}\bar{\theta},
\end{equation}
which favors localization at the $\bar{\theta}\simeq\pi/2$ plane. In this
expression $\Im\omega(\ell)$ corresponds to the imaginary
part of the eigenfrequency $\omega_{1,\ell,\ell,0}$, which is assumed
positive for the unstable mode. 

In the case where the dual theory lives in flat coordinates, we can in
principle just apply the transformation \eqref{eq:flatcoordchange} to
translate from the global results above. However, our analysis in
Sec. \ref{sec:scalarperturbations} was made by assuming $a_1>a_2$,
which in global coordinates is not a restriction because one can
interchange the azimuthal variables $\bar{\phi}_1$ and
$\bar{\phi}_2$. When we transform to flat coordinates, we choose one
of them to map to the coordinate $\hat{\phi}$, and we have to
consider the two choices separately.

For the case $a_1>a_2$, we have $\hat{\phi}=\bar{\phi}_2$ as in
Sec. \ref{sec:scalarperturbations}, and the energy profile for the
fluctuation reads 
\begin{equation}
  \bar{\rho}_{\mathrm{fluc.}}\propto
  e^{2\Im \omega(\ell)\bar{t}}\left(
    \frac{(1+\tfrac{1}{4}(\hat{t}^2-\hat{r}^2))^2+\hat{r}^2\cos^2\hat{\theta}}{
      (1+\tfrac{1}{4}(\hat{t}+\hat{r})^2)(1+\tfrac{1}{4}(\hat{t}-\hat{r})^2)}
  \right)^{\ell/2}
\end{equation}
where
\begin{equation}
  \bar{t} =
  \arctan\frac{\hat{t}+\hat{r}}{2}+\arctan\frac{\hat{t}-\hat{r}}{2}.
  \label{eq:thisistbar}
\end{equation}
The case $a_1<a_2$, can be realized by setting
$\hat{\phi}=\bar{\phi}_1$, and the coordinate transformation changes from 
\eqref{eq:flatcoordchange} to 
\begin{equation}
  \sin\bar{\theta}=\sin\hat{\chi}\sin\hat{\theta},
\end{equation}
which corresponds to interchanging $a_1\leftrightarrow a_2$ in
the profile \eqref{eq:flatdelta}. The energy dissipated in the
perturbation mode is now
\begin{equation}
  \bar{\rho}_{\mathrm{fluc.}}\propto
  e^{2\Im \omega(\ell)\bar{t}}\left(
    \frac{\hat{r}^2\sin^2\hat{\theta}}{
      (1+\tfrac{1}{4}(\hat{t}+\hat{r})^2)(1+\tfrac{1}{4}(\hat{t}-\hat{r})^2)}
  \right)^{\ell/2},
  \label{eq:decaya2a1}
\end{equation}
with $\bar{t}$ given by \eqref{eq:thisistbar}.

We note that, with the second choice, there is a tendency for the decay
to occur in the $\hat{x}-\hat{y}$ plane ($\sin^2\hat{\theta}\simeq
1$), following the maximum of $\bar{\rho}_{\mathrm{fluc.}}$. As we
discussed above, the specific value of $\ell$ which will be
preferred by the decay will depend on the details of the state and the
coupling. In Fig. \ref{fig:qnm123ell} we see that, even though the
maximum of the imaginary part of the eigenfrequency decreases with $\ell$,
larger values of $r_+$ tend to favor larger values of $\ell$ for the decay.
This leads us to infer that, at low temperatures, the localization
effect of the decay products resulting from \eqref{eq:decaya2a1}
increases with the mass ${\cal M}$. We also point out the difference
between the time dependence of the decay modes depending whether we
consider global coordinates $\mathbb{R}\times S^3$ (bar
coordinates) or conformally flat coordinates $\mathbb{R}^{3,1}$
(hatted coordinates). Whereas in global coordinates the fluctuations
grow exponentially without bound, at least in the linear analysis we
do here, the growth in conformally flat coordinates is capped as
$\hat{t}\rightarrow \infty$ due to \eqref{eq:thisistbar}.

\section{Discussion}
\label{sec:discussion}

In this paper we analyzed the holographic aspects of the generically
rotating Kerr-AdS$_5$ black hole in Lorentzian signature, focussing
on the low-temperature limit $T_+\simeq 0$ and small black holes
$r_+\ll 1$. We reviewed the definition of asymptotic charges,
discussed the geometrical meaning of the regularization involved in 
the definition of mass, and determined the holographic stress-energy
tensor associated to the black hole in the dual field theory -- whose
thermal state is characterized by non-zero vacuum expectation values for
all generators of the Weyl subgroup of the conformal group
$\mathrm{SO}(4,2)$. Finally, we computed the first correction to the
boundary-to-boundary scalar propagator due to the presence of the
black hole and found that it makes scalar perturbations with dimension close
to the marginal value ($\Delta=4$) irrelevant, thus suggesting that
such perturbations should not substantially change the 
fate of the background as the theory flows to the IR.

In order to further analyze this question, we set out to treat scalar
perturbations by means of the exact solution constructed in
\cite{Barragan-Amado:2018pxh}, adapted to the low-temperature
limit. We found instabilities in the QNMs for all values of the
angular eigenvalue  $\ell>0$, as anticipated by \cite{Hawking:2000vt},
using thermodynamical arguments. For $\ell=0$, the first order
correction in the temperature has an enhancement effect which
increases the decay time of the perturbation, but it is not strong
enough to induce instabilities. We have also studied
numerically the dependence of the $\ell>0$ instabilities on the black
hole parameters as we vary $\ell$. These instabilities of AdS space
are expected from general arguments \cite{Green:2015kur}, and we
attempted a holographic interpretation by studying the instabilities
of the thermal field theory state corresponding to the black hole and
the spatial dependence of its decay products. We found qualitative
hints that the ejecta tend to collimate at the axis of highest
rotation as the black hole mass increases.

We have also shown that the general procedure of
\cite{Barragan-Amado:2018pxh}, which solved for the non-local boundary
conditions related to QNMs in terms of monodromy data of the Heun
differential equations involved, can be used effectively to find QNMs
in the zero temperature limit. This limit led us to consider the
monodromy parameters to be determined by the Painlevé V transcendent,
obtained from the confluence limit of the Painlevé VI. The importance
of the expansion of the Painlevé transcendents in terms of $c=1$
Virasoro conformal blocks, as studied in \cite{Lisovyy:2018mnj},
provides us with the same interesting factorization of the
\textit{four-dimensional} conformal blocks in terms of
(semi-classical) \textit{two-dimensional} ones, as first anticipated by
\cite{Dolan:2003hv}. The new ingredient here is that the
two-dimensional conformal blocks are of the irregular type, as also 
appeared in the black hole perturbation context in
\cite{CarneirodaCunha:2019tia}. The semiclassical conformal blocks
associated to the radial perturbations seem to arise from a unitary
theory, and the Vertex operators associated to the inner and outer
horizons can be interpreted as thermal states, with their degeneracy
equating the entropy of the scalar perturbation with quantum numbers
given by $\omega,m_1$ and $m_2$ as it is absorbed by the black
hole. Finally, the study of QNMs at small values of $r_+$ showed the
relevance of short Virasoro representations for the intermediate
states of the conformal blocks, again anticipated in a different
context by \cite{Fitzpatrick:2015foa}. 

The motivation for considering low-temperature black holes stemmed
from the necessity for pushing the hydrodynamical analogy of
\cite{Bhattacharyya:2007vs}, if we want to understand the holographic
interpretation of instabilities in AdS space. We have seen that the
phase diagram of these states in the dual theory has a richer set 
of associated phenomena than what is expected from the ``naive''
near-UV considerations, with instabilities in the higher modes and
collimation of decay products. Tools usually used in holographic
renormalization-group flow are not really suitable to determine the
fate of the state as one flows to the infrared. We can use the same
methods proposed here to study what happens at larger values of $r_+$,
but the appearance of instabilities as well as fragmentation issues
\cite{Emparan:2003sy} -- the latter absent here -- can further
contribute to complicate the issue of determining the fate of
the decay process.

Finally, we stress that the isomonodromic method used here for the
numerical analysis overcomes a major hurdle arising in usual Frobenius
matching. At low temperatures, the latter method suffers from large
spatial oscillations of the modes near the outer horizon $r_+$. The
isomonodromic method has no such hindrance. Moreover, our
investigation also shows that the mode expansion used in Frobenius
matching is not suitable to study the fundamental mode. Indeed, in the
finite temperature case, we have shown that the monodromy parameters,
and thus the actual value of the QNM frequency, receive
contributions from all levels in the Frobenius expansion. These
contributions can be interpreted, in the two-dimensional conformal
block attribution, as coming from all descendants of the CFT primaries
associated to the intermediate states. We verified that these
contributions can be resummed in terms of the generating function of
the Catalan numbers. Also, short representations of the CFT Virasoro
algebra are relevant for the calculation of QNMs at higher levels. We defer
the study of other interesting black hole limits, as well as higher
spin perturbations \cite{Barragan-Amado:2020pad} to future work.

\section*{Acknowledgements}

We apologize for any omission in this important and long
standing field of research. The authors would like to thank Dmitry
Melnikov, Monica Guica, Oleg Lisovyy for comments and suggestions
along the way. Finally, we thank the referee for pointing out the
necessity of further comparison between \eqref{eq:fundomega} and our
numerical results, motivating a revision of the fundamental mode
frequency in a previous version of this paper.

\appendix

\section{Asymptotic charges of AdS spaces}
\label{sec:adsmass}

Let the asymptotic metric be given by
\begin{equation}
  g_{ab}=\frac{1}{\tilde{\Omega}^2}\tilde{g}_{ab}+h_{ab}
\end{equation}
where $h_{ab}$ is understood to be small with respect to the first
term in the $\tilde{\Omega}\rightarrow 0$ limit, in a manner that will become
precise at the end of the Appendix. As we will see, for asymptotically
AdS spaces, $\tilde{\Omega}$ has spacelike gradient and vanishes at the
conformal boundary. The boundary structure is dependent on the
particular choice of $\tilde{\Omega}$, and the two choices considered
in this paper are $\tilde{g}_{ab}$ is either the flat
metric (Poincaré patch), called $\hat{g}_{ab}$ in the main text, or the
$\mathbb{R}\times S^3$ metric (global coordinates), referred to by
$\bar{g}_{ab}$ in the main text. Let us introduce the covariant derivates $\nabla_a$,
associated with $g_{ab}$ and $\tilde{\nabla}_a$, associated with
$\tilde{g}_{ab}$. If $\chi^a$ is a vector field, we can use Leibniz
rule to show that the difference
$\nabla_a\chi^b-\tilde{\nabla}_a\chi^b$ is a linear operator on 
$\chi^a$, therefore, for any two covariant derivatives, we have
\begin{equation}
  \nabla_a\chi^b=\tilde{\nabla}_a\chi^b+{C^b}_{ac}\chi^c,
  \label{eq:tildeconnection}
\end{equation}
with the difference between the connections given by the conditions
that $\nabla_a$ is compatible with $g_{ab}$ and $\tilde{\nabla}_a$ with
$\tilde{g}_{ab}$. To first order in $h_{ab}$:
\begin{align}
  {C^b}_{ac} & = \frac{1}{2} g^{bd}(\tilde{\nabla}_a g_{dc}+\tilde{\nabla}_c
  g_{ad}-\tilde{\nabla}_d g_{ac}) \\
  & =\frac{1}{2}\tilde{\Omega}^2(\tilde{g}^{bd}-\tilde{\Omega}^2h^{bd} )(
  \tilde{\nabla}_a(\frac{1}{\tilde{\Omega}^2}\tilde{g}_{dc}+h_{dc})+
  \tilde{\nabla}_c(\frac{1}{\tilde{\Omega}^2}\tilde{g}_{ad}+h_{ad})-
    \tilde{\nabla}_d(\frac{1}{\tilde{\Omega}^2}\tilde{g}_{ac}+h_{ac})),
    \label{eq:tildelevicivita}
\end{align}
with indices in the right hand side now and hereafter raised with
$\tilde{g}^{ab}$, so that
$h^{ab}\equiv\tilde{g}^{ac}h_{cd}\tilde{g}^{db}$. Expanding up to quadratic
terms in $h_{ab}$, 
\begin{multline}
  {C^b}_{ac} = -\frac{1}{\tilde{\Omega}}(\delta_c^b\tilde{\nabla}_a\tilde{\Omega}+
  \delta_a^b\tilde{\nabla}_c\tilde{\Omega}-\tilde{g}_{ac}\tilde{\nabla}^b\tilde{\Omega})
  +\tilde{\Omega}
  (h_c^b\tilde{\nabla}_a\tilde{\Omega}+h_a^b \tilde{\nabla}_c\tilde{\Omega}-
  \tilde{g}_{ac}h^{bd}\tilde{\nabla}_d\tilde{\Omega} ) \\
  + \tfrac{1}{2}\tilde{\Omega}^2(\tilde{\nabla}_ah_c^b+
  \tilde{\nabla}_ch_a^b-\tilde{\nabla}^bh_{ac}).
\end{multline}

Conserved quantities associated with vector fields $\xi^a$ are
defined as the integral of the $(d-2)$-form 
\begin{equation}
  Q[\xi]_{a_1\ldots
    a_{d-2}}=-\frac{1}{2\kappa}\epsilon[g]_{aba_1\ldots
    a_{d-2}}g^{ac}\nabla_c\xi^b, 
\end{equation}
with $\epsilon[g]$ the volume form associated with
$g_{ab}$\footnote{For convenience, a term equal to the volume of the
  $d-2$ sphere dividing $Q[\xi]$ is omitted.}. The
contravariant derivative can also be computed to first order in 
$h_{ab}$:
\begin{equation}
  g^{ac}\nabla_c\xi^b=\tilde{\Omega}^2(\tilde{g}^{ac}-\tilde{\Omega}^2 h^{ac})(
  \tilde{\nabla}_c\xi^b+{C^b}_{cd}\xi^d)
\end{equation}
or
\begin{multline}
g^{ac}{C^b}_{cd} = -\tilde{\Omega}(\delta_d^b\tilde{\nabla}^a\tilde{\Omega}+
  \tilde{g}^{ab}\tilde{\nabla}_d\tilde{\Omega}-\delta^a_d\tilde{\nabla}^b\tilde{\Omega})
  +\tilde{\Omega}^3
  (h^b_d\tilde{\nabla}^a\tilde{\Omega}+h^{ab} \tilde{\nabla}_d\tilde{\Omega}-
  h^a_d\tilde{\nabla}^b\tilde{\Omega}) \\
  +\tilde{\Omega}^3(\delta_d^bh^{ac}\tilde{\nabla}_c\tilde{\Omega} +
  h^{ab}\tilde{\nabla}_d\tilde{\Omega} -\delta_d^ah^{bc}
  \tilde{\nabla}_c\tilde{\Omega}) +
  \tfrac{1}{2}\tilde{\Omega}^4(\tilde{\nabla}^ah_d^b+ \tilde{\nabla}_dh^{ab}-
  \tilde{\nabla}^bh_d^a), 
\end{multline}

Now, the vector fields we are interested in are Killing vector fields
of the induced metric at the conformal boundary $\tilde{\Omega}=0$. These will
satisfy: 
\begin{equation}
  \xi^d\tilde{\nabla}_d\tilde{\Omega} = 0,\quad\quad
  \tilde{\nabla}^a\xi^b+\tilde{\nabla}^b\xi^a=0,
\end{equation}
denoting the vanishing Lie derivatives with respect to $\xi^a$ of
$\tilde{\Omega}$ and $\tilde{g}^{ab}$, respectively. Note that we are
assuming that $\xi^a$ and $\tilde{\nabla}^a\tilde{\Omega}$ are orthogonal,
which is suitable for the asymptotically AdS case.

Wrapping it all up,
\begin{multline}
  g^{ac}\nabla_c\xi^b = 2\tilde{\Omega} \xi^{[a}\tilde{\nabla}^{b]}\tilde{\Omega}
  +\tilde{\Omega}^2\tilde{\nabla}^{[a}\xi^{b]}
  -2\tilde{\Omega}^3 (\xi^dh_d^{[a}\tilde{\nabla}^{b]}\tilde{\Omega}
  +\xi^{[a}h^{b]c}\tilde{\nabla}_c\tilde{\Omega}) \\
  +\tilde{\Omega}^4(\xi^d\tilde{\nabla}^{[a}h_d^{b]}-
  h^{d[a}\tilde{\nabla}_d\xi^{b]}+\tfrac{1}{2}\mathsterling_\xi h^{ab})+{\cal O}(h^2)
\end{multline}
where $\mathsterling_\xi h^{ab}$ is the Lie derivative of $h^{ab}$ with
respect to $\xi^a$. The volume form is readily computed:
\begin{equation}
  {\bm{\epsilon}}[g] = \frac{1}{\tilde{\Omega}^d}{\bm{\epsilon}}
  [\tilde{g}](1+\tfrac{1}{2}\tilde{\Omega}^2 h_a^a+ 
  {\cal O}(h^2))
\end{equation}
and so the charge density can be expanded
\begin{multline}
  Q[\xi]_{a_1\ldots a_{d-2}}=-\frac{1}{2\kappa}
  \epsilon[\tilde{g}]_{ab a_1\ldots a_{d-2}}\frac{1}{\tilde{\Omega}^{d-1}}(
    2\xi^{[a}\tilde{\nabla}^{b]}\tilde{\Omega}
  +\tilde{\Omega}\tilde{\nabla}^{[a}\xi^{b]} \\
  -2\tilde{\Omega}^2 (\xi^dh_d^{[a}\tilde{\nabla}^{b]}\tilde{\Omega}
  +\xi^{[a}h^{b]c}\tilde{\nabla}_c\tilde{\Omega} - \tfrac{1}{2}
  h_c^c\xi^{[a}\tilde{\nabla}^{b]}\tilde{\Omega} ) \\ 
  +\tilde{\Omega}^3(\xi^d\tilde{\nabla}^{[a}h_d^{b]}-
  h^{d[a}\tilde{\nabla}_d\xi^{b]}+\tfrac{1}{2}h_c^c
  \tilde{\nabla}^{[a}\xi^{b]})+{\cal O}(h^2))
  \label{eq:firstq}
\end{multline}
By integrating ${\mathbf Q}[\xi]$ in the conformal boundary
$\tilde{\Omega}\rightarrow 0$ limit we can find the conserved
quantity. Supposing a falloff behavior of the stress-energy tensor,
the Einstein equation for negative cosmological constant
(and $\ell_{AdS}=1$) near infinity is written as:
\begin{equation}
  R_{ac}=-(d-1){g_{ac}},
\end{equation}
which, when written in terms of $\tilde{g}_{ab}$, yields, up to terms
of order $h_{ab}$:
\begin{equation}
  \tilde{R}_{ac}+(d-2)\frac{\tilde{\nabla}_a\tilde{\nabla}_c\tilde{\Omega}}{\tilde{\Omega}}
  -(d-1)\tilde{g}_{ac}\tilde{g}^{bd}\frac{\tilde{\nabla}_b\tilde{\Omega}
    \tilde{\nabla}_d\tilde{\Omega}}{
    \tilde{\Omega}^2}+\tilde{g}_{ac}\tilde{g}^{bd}\frac{\tilde{\nabla}_b
    \tilde{\nabla}_d\tilde{\Omega}}{\tilde{\Omega}} =
  -(d-1)\frac{\tilde{g}_{ac}}{\tilde{\Omega}^2}.
\end{equation}
Requiring that the term proportional to $\tilde{\Omega}^{-2}$ vanishes at the
boundary implies that $n^a=\tilde{g}^{ab}\tilde{\nabla}_b\tilde{\Omega}$ is
normalized space-like according to $\tilde{g}_{ab}$ in the
$\tilde{\Omega}\rightarrow 0$ limit. We note, however, that the order ${\cal
  O}(\tilde{\Omega}^{-1})$ term will set the (conformal, Ricci)
geometry at infinity through 
\begin{equation}
  \tilde{R}_{ac}-\frac{1}{2(d-1)}\tilde{g}_{ac}\tilde{R}=-(d-2)\frac{1}{\tilde{\Omega}}
  \tilde{\nabla}_a\tilde{\nabla}_c\tilde{\Omega}.
  \label{eq:confstruct}
\end{equation}
Therefore the conformal geometry chosen for the four-dimensional
manifold at infinity will influence the asymptotics of $\tilde{\Omega}$, and
thus the actual value of the asymptotic charges.

We may consider now two cases for $\mathbf{Q}[\xi]$, depending on $\xi^a$:
\begin{enumerate}
  \item $\xi^a$ is the time translation operator, and therefore
    orthogonal to the surface of integration used to define the
    conserved quantity;
  \item $\xi^a$ generates rotations, so it is tangent to the surface
    of integration;
\end{enumerate}
In the first case, the term $\xi^{[a}\tilde{\nabla}^{b]}\tilde{\Omega}$ is a
bi-vector normal to the surface of integration 
and $\mathbf{Q}[\xi]$ diverges in the $\tilde{\Omega}\rightarrow 0$
limit.  One usually deals with this divergence by adding a boundary
(Gibbons-Hawking-York) term to the action to regularize it, although
the procedure of fixing the conformal structure, much in the spirit of
what has been done in this calculation, was defended in
\cite{Witten:2018lgb}. For the perturbed conformal metric, the vector
$\tilde{\nabla}^a\tilde{\Omega}$ approaches the unit vector at spatial
infinity $n^a$ by:
\begin{equation}
  n^a=(1+N)^{-1/2}\tilde{\nabla}^a\tilde{\Omega},\quad\quad
  N=\tilde{\Omega}^2h_{ab}\tilde{\nabla}^a
  \tilde{\Omega}\tilde{\nabla}^b\tilde{\Omega}.
\end{equation}

The second term in brackets in the first line of \eqref{eq:firstq} is
$\tilde{\Omega}\tilde{\nabla}^{[a}\xi^{b]}$ which we 
will also take to vanish, either because $\xi^a$ is covariantly
constant with respect to $\tilde{g}_{ab}$, as in the case of time
translation, or because the 
derivative is tangent to the surface of integration, as in the case of
rotations. The next terms will converge in the
$\tilde{\Omega}\rightarrow 0$ limit if
$h_{ab}=\tilde{\Omega}^{d-3}\gamma_{ab}$. Substituting in \eqref{eq:firstq}
and taking the limit we have:
\begin{equation}
  Q[\xi]_{a_1\ldots a_{d-2}}=\frac{1}{\kappa}
  \epsilon[\tilde{g}]_{aba_1 \ldots
    a_{d-2}}((d-1)n^{[a}\gamma^{b]}_c\xi^c-2\xi^{[a}\gamma^{b]}_cn^c
  +(\gamma^c_c+N)\xi^{[a}n^{b]})+{\cal O}(\tilde{\Omega}).
  \label{eq:secondq}
\end{equation}
One can then conclude that the value for generic charges in
asymptotically AdS spaces stems not only from the conformal
structure encoded in $\tilde{g}$, but also from the notion of
``normalized direction in RG flow'', encoded by the normalization of
$n^a$.

\section{The Fredholm determinant formulation of the Painlevé VI
  transcendent}
\label{sec:painleve}

This is a review of the Fredholm determinant formulation of the
Painlevé VI, borrowing heavily from \cite{Gavrylenko:2016zlf}
\begin{equation}
  \tau(t)=\mathrm{const}\cdot
  t^{\frac{1}{4}(\sigma^2-\theta_0^2-\theta_t^2)}
  (1-t)^{-\frac{1}{2}\theta_t\theta_1}\det(\mathbbold{1}-\mathsf{A}
  \Phi(t)\mathsf{D}\Phi(t)^{-1}),
\label{eq:fredholmexpansion}
\end{equation}
where the Plemelj operators $\mathsf{A},\mathsf{D}$ act on the space of pairs of 
square-integrable functions defined on ${\cal C}$, a circle on the
complex plane with radius $R<1$:
\begin{equation}
  (\mathsf{A}g)(z)=\oint_{\cal C} \frac{dz'}{2\pi i}A(z,z')g(z'),\quad
  (\mathsf{D}g)(z)=\oint_{\cal C} \frac{dz'}{2\pi i}D(z,z')g(z'),\quad
  g(z)=\begin{pmatrix}
    f_+(z) \\
    f_-(z)
  \end{pmatrix},
  \label{eq:fredholmad}
\end{equation}
with kernels given, for $|t|<R$, explicitly by 
\begin{equation}
  \begin{gathered}
    A(z,z')=\frac{\Psi(\sigma,\theta_1,\theta_\infty;
      z)\Psi^{-1}(\sigma,\theta_1,\theta_\infty;z')-\mathbbold{1}}{z-z'},\\ 
    D(z,z')=\frac{\mathbbold{1}-\Psi(-\sigma,\theta_t,\theta_0;t/z)
      \Psi^{-1}(-\sigma,\theta_t,\theta_0;t/z')}{z-z'}.
  \end{gathered}
\end{equation}
The operators $\mathsf{A}$ and $\mathsf{D}$ can be thought of as
projecting the analytic and principal parts of a generic function
defined on the circle ${\cal C}$ into the space generated by functions with
definite monodromy. The parametrix $\Psi$ and the  ``gluing''
matrix $\Phi$ are 
\begin{equation}
  \resizebox{.90\hsize}{!}{$\displaystyle
    \Psi(\alpha_1,\alpha_2,\alpha_3;z) = \begin{pmatrix}
      \phi(\alpha_1,\alpha_2,\alpha_3;z) &
      \chi(\alpha_1,\alpha_2,\alpha_3;z) \\
      \chi(-\alpha_1,\alpha_2,\alpha_3;z) &
      \phi(-\alpha_1,\alpha_2,\alpha_3;z)
    \end{pmatrix},\quad
    \Phi(\kappa,\sigma;t)=\begin{pmatrix}
      t^{-\sigma/2}\kappa^{-1/2} & 0 \\
      0 & t^{\sigma/2}\kappa^{1/2}
    \end{pmatrix}$},
  \label{eq:parametrixphi}
\end{equation}
with $\phi$ and $\chi$ given in terms of Gauss' hypergeometric
function -- note the overall minus sign with respect to the conventions in
\cite{Gavrylenko:2016zlf}:
\begin{equation}
  \resizebox{.95\hsize}{!}{$\displaystyle 
  \begin{gathered}
    \phi(\alpha_1,\alpha_2,\alpha_3;z) = {_2F_1}(
    \tfrac{1}{2}(\alpha_1-\alpha_2+\alpha_3),\tfrac{1}{2}(\alpha_1-\alpha_2-\alpha_3);
    \alpha_1;z) \\
    \chi(\alpha_1,\alpha_2,\alpha_3;z) =
    \frac{\alpha_3^2-(\alpha_1-\alpha_2)^2}{4\alpha_1(1+\alpha_1)}
      z\,{_2F_1}(
      1+\tfrac{1}{2}(\alpha_1-\alpha_2+\alpha_3),
      1+\tfrac{1}{2}(\alpha_1-\alpha_2-\alpha_3);
      2+\alpha_1;z).
    \end{gathered}
     $}
\end{equation}
Finally, $\kappa$ is a known function of the monodromy
parameters:
\begin{multline}
  \kappa=s\frac{\Gamma^2(1-\sigma)}{\Gamma^2(1+\sigma)}
  \frac{\Gamma(1+\tfrac{1}{2}(\theta_t+\theta_0+\sigma))
    \Gamma(1+\tfrac{1}{2}(\theta_t-\theta_0+\sigma))}{
    \Gamma(1+\tfrac{1}{2}(\theta_t+\theta_0-\sigma))
    \Gamma(1+\tfrac{1}{2}(\theta_t-\theta_0-\sigma))} \times \\
    \frac{\Gamma(1+\tfrac{1}{2}(\theta_1+\theta_\infty+\sigma))
    \Gamma(1+\tfrac{1}{2}(\theta_1-\theta_\infty+\sigma))}{
    \Gamma(1+\tfrac{1}{2}(\theta_1+\theta_\infty-\sigma))
    \Gamma(1+\tfrac{1}{2}(\theta_1-\theta_\infty-\sigma))},
\label{eq:tildes}
\end{multline}
and, finally, the parameter $s$ is given in terms of the monodromy
parameters $\{\sigma_{0t},\sigma_{t1}\}$:
\begin{equation}
  s=\frac{(w_{1t}-2p_{1t}-p_{0t}p_{01}) - (w_{01}-2p_{01}-p_{0t}p_{1t})
    \exp({\pi i \sigma_{0t}})}{(2\cos \pi
    (\theta_t-\sigma_{0t})-p_0)(2\cos \pi
    (\theta_1-\sigma_{0t})-p_{\infty})},
  \label{eq:ess}
\end{equation}
where
\begin{equation}
\begin{gathered}
  p_i=2\cos\pi\theta_i,\quad  p_{ij}=2\cos\pi\sigma_{ij}, \\
  w_{0t}=p_0p_t+p_1p_{\infty},\quad w_{1t}=p_1p_t+p_0p_{\infty},\quad
  w_{01}=p_0p_1+p_tp_{\infty}.
\end{gathered}
\end{equation}

The expansion for the tau function at small $t$ can be computed
directly from its definition:
\begin{multline}
  \tau(t) =
  Ct^{\frac{1}{4}(\sigma^2-\theta_0^2-\theta_t^2)}(1-t)^{\frac{1}{2}\theta_1\theta_t}
  \left(1+\left(\frac{\theta_1\theta_t}{2}+
    \frac{(\theta_0^2-\theta_t^2-\sigma^2)(\theta_\infty^2-
      \theta_1^2-\sigma^2)}{8\sigma^2}\right)t\right.\\
  -\frac{(\theta_0^2-(\theta_t-\sigma)^2)
      (\theta_\infty^2-(\theta_1-\sigma)^2)}{
      16\sigma^2(1+\sigma)^2}\kappa t^{1+\sigma}\\
    \left. -\frac{(\theta_0^2-(\theta_t+\sigma)^2)
      (\theta_\infty^2-(\theta_1+\sigma)^2)}{
      16\sigma^2(1-\sigma)^2}\kappa^{-1}t^{1-\sigma}
  +\ldots\right).
\label{eq:taufunctionexpansion}
\end{multline}

The accessory parameter $K_0$, defined
by \eqref{eq:monoconditions} as essentially the logarithm derivative
of $\tau$, has an analogous expansion for small $t$
and can be seen in \cite{Barragan-Amado:2018pxh}.


\providecommand{\href}[2]{#2}\begingroup\raggedright\endgroup

\end{document}